\documentclass[12pt]{article}

\usepackage{amsmath}
\usepackage{epsf}

\DeclareSymbolFont{boldletters}{OML}{cmm} {b}{it}
\DeclareSymbolFontAlphabet{\mathbit}{boldletters}
\DeclareMathSymbol{\alpha}{\mathalpha}{letters}{"0B}
\DeclareMathSymbol{\beta}{\mathalpha}{letters}{"0C}
\DeclareMathSymbol{\gamma}{\mathalpha}{letters}{"0D}
\DeclareMathSymbol{\delta}{\mathalpha}{letters}{"0E}
\DeclareMathSymbol{\epsilon}{\mathalpha}{letters}{"0F}
\DeclareMathSymbol{\zeta}{\mathalpha}{letters}{"10}
\DeclareMathSymbol{\eta}{\mathalpha}{letters}{"11}
\DeclareMathSymbol{\theta}{\mathalpha}{letters}{"12}
\DeclareMathSymbol{\iota}{\mathalpha}{letters}{"13}
\DeclareMathSymbol{\kappa}{\mathalpha}{letters}{"14}
\DeclareMathSymbol{\lambda}{\mathalpha}{letters}{"15}
\DeclareMathSymbol{\mu}{\mathalpha}{letters}{"16}
\DeclareMathSymbol{\nu}{\mathalpha}{letters}{"17}
\DeclareMathSymbol{\xi}{\mathalpha}{letters}{"18}
\DeclareMathSymbol{\pi}{\mathalpha}{letters}{"19}
\DeclareMathSymbol{\rho}{\mathalpha}{letters}{"1A}
\DeclareMathSymbol{\sigma}{\mathalpha}{letters}{"1B}
\DeclareMathSymbol{\tau}{\mathalpha}{letters}{"1C}
\DeclareMathSymbol{\upsilon}{\mathalpha}{letters}{"1D}
\DeclareMathSymbol{\phi}{\mathalpha}{letters}{"1E}
\DeclareMathSymbol{\chi}{\mathalpha}{letters}{"1F}
\DeclareMathSymbol{\psi}{\mathalpha}{letters}{"20}
\DeclareMathSymbol{\omega}{\mathalpha}{letters}{"21}
\DeclareMathSymbol{\varepsilon}{\mathalpha}{letters}{"22}
\DeclareMathSymbol{\vartheta}{\mathalpha}{letters}{"23}
\DeclareMathSymbol{\varpi}{\mathalpha}{letters}{"24}
\DeclareMathSymbol{\varrho}{\mathalpha}{letters}{"25}
\DeclareMathSymbol{\varsigma}{\mathalpha}{letters}{"26}
\DeclareMathSymbol{\varphi}{\mathalpha}{letters}{"27}
\DeclareMathSymbol{\Gamma}{\mathalpha}{letters}{"00}
\DeclareMathSymbol{\Delta}{\mathalpha}{letters}{"01}
\DeclareMathSymbol{\Theta}{\mathalpha}{letters}{"02}
\DeclareMathSymbol{\Lambda}{\mathalpha}{letters}{"03}
\DeclareMathSymbol{\Xi}{\mathalpha}{letters}{"04}
\DeclareMathSymbol{\Pi}{\mathalpha}{letters}{"05}
\DeclareMathSymbol{\Sigma}{\mathalpha}{letters}{"06}
\DeclareMathSymbol{\Upsilon}{\mathalpha}{letters}{"07}
\DeclareMathSymbol{\Phi}{\mathalpha}{letters}{"08}
\DeclareMathSymbol{\Psi}{\mathalpha}{letters}{"09}
\DeclareMathSymbol{\Omega}{\mathalpha}{letters}{"0A}
\newcommand{\mbit}[1]{{\mathbit#1}}

\newcommand{\dbox}{\,\framebox(7,7)[t]{}\,}
 
\begin{document}

\title{
\begin{flushright}
\small{
Ehime-th-5 \\
KYUSHU-HET-90}
\end{flushright}
The Auxiliary Field Method in Quantum Mechanical Four-Fermi Models \\
-- {\Large A Study Toward Chiral Condensation in QED} --}

\author{Taro~Kashiwa\thanks{kashiwa@phys.sci.ehime-u.ac.jp}\\  
Department of Physics, Ehime University, Matsuyama 790-8577, Japan \\
\\
Tomohiko~Sakaguchi\thanks{tomohiko@higgs.phys.kyushu-u.ac.jp}
\\
Department of Physics, Kyushu University, Fukuoka 812-8581, Japan\\\\}

\date{\today}

\maketitle

\abstract{A study for checking validity of the auxiliary field method (AFM) is made in quantum mechanical four-fermi models which act as a prototype of models for chiral symmetry breaking in Quantum Electrodynamics. It has been shown that AFM, defined by an insertion of Gaussian identity to path integral formulas and by the loop expansion, becomes more accurate when taking higher order terms into account under the bosonic model with a quartic coupling in 0- and 1-dimensions as well as the model with a four-fermi interaction in 0-dimension. The case is also confirmed in terms of two models with the four-fermi interaction among $N$ species in 1-dimension (the quantum mechanical four-fermi models): higher order corrections lead us toward the exact energy of the ground state. It is found that the second model belongs to a WKB-exact class that has no higher order corrections other than the lowest correction. Discussions are also made for unreliability on the continuous time representation of path integration and for a new model of QED as a suitable probe for chiral symmetry breaking.}

\maketitle\thispagestyle{empty}
\newpage

\section{Introduction}

Chiral condensation, characterized by that the expectation value of fermi field operators, $\langle \overline{\psi }\psi \rangle $, becomes nonzero, is a nonperturbative phenomenon: in relativistic field theory a model with four-fermi interactions proposed by Nambu--Jona-Lasinio \cite{rf:NJL} has been served as a simple example exhibiting chiral symmetry breaking. The Lagrangian reads 
\begin{eqnarray}
{\cal L}=  - \overline{\psi}\gamma_\mu \partial_\mu  \psi +  \frac{g^2}{2}\left\{  \left(  \overline{\psi} \psi  \right)^2 -  \left(  \overline{\psi}\gamma_5  \psi  \right)^2  \right\}  \ . \label{NJLLagrangian}
\end{eqnarray}
Originally the mean field had been adopted to deduce the gap equation \cite{rf:NJL}, but later a more powerful prescription has been dispensed by Gross-Neveu then Kugo and Kikkawa \cite{rf:GNKK}: the method is evolved by inserting an identity in terms of the Gaussian integration with respect to a fictitious field called an auxiliary field or the Hubbard-Stratonovich field \cite{rf:FRAD} into a path integral expression of Eq.~(\ref{NJLLagrangian}). (As for the path integral expression see \cite{rf:KOS}.) The prescription combined with the loop expansion \cite{rf:KOS1} is simple and transparent and is called the auxiliary field method (AFM) \cite{rf:KOS2}. (Any approximation schemes cannot spread out widely, unless those are simple and transparent. The Feynman graph technique is a typical example.) If the number of fermion $N$ is large enough quantum corrections of the auxiliary field is negligible. The gap equation obtained under this assumption exhibits chiral symmetry breaking. However if $N$ is small, quantum corrections, that is, higher loop contributions may change the phase structure \cite{rf:KS1}.

A question would, then, be raised that how effective is AFM in quantum mechanical systems? Accordingly  we have investigated the role of auxiliary fields in the bosonic case with a quartic interaction in 0- and 1-dimension \cite{rf:kashiwa} and in 0-dimensional four-fermi case \cite{rf:KS2}. These models can be solved exactly or numerically so that we can check how accurate is the result of AFM. We find that in the bosonic case AFM does work excellently when we take higher loops into account. (Of course, the loop expansion is the asymptotic expansion so that we should stop considering higher loops somewhere.)

In order to illustrate AFM, an outline of the 0-dimensional fermionic
model is given~\cite{rf:KS2}: the target quantity is the ``partition
function'' $Z$ (although in 0 dimension),
\begin{eqnarray}
   Z &=& \int d^N \mbit{\xi} d^N \mbit{\xi}^*  \exp \left[
{-\omega(\mbit{\xi}^* \cdot \mbit{\xi}) 
+\frac{\lambda^2}{2N}}(\mbit{\xi}^*  \cdot \mbit{\xi})^2 \right] \ ,
\label{gras}
\end{eqnarray}
where
\begin{eqnarray}
   d^N \mbit{\xi} \equiv  d \xi_1 \cdot \cdot \cdot d \xi_N \ , \quad
   d^N \mbit{\xi}^* \equiv  d \xi^*_N \cdot \cdot \cdot d \xi^*_1 \ , \quad
   (\mbit{\xi}^*  \cdot \mbit{\xi}) \equiv  \sum_{i=1}^N \xi^*_i  \xi_i \ ,
\end{eqnarray}
and the coupling constant $\lambda^2$ is supposed real. We have introduced $2N$-Grassmann variables and the notation is followed from the textbook of ref.~\cite{rf:KOS3}. Introduce an auxiliary field, $y$, to kill the $(\mbit{\xi}^* \cdot \mbit{\xi} )^2$ term in Eq.~(\ref{gras}), which can be realized by inserting the identity
\begin{eqnarray}
  1 = \int_{-\infty}^{\infty} \frac{d y}{\sqrt{2\pi}}
     \exp \left[ -\frac{1}{2} \left\{   y + \frac{\lambda}{\sqrt{N}}
     (\mbit{\xi}^*  \cdot \mbit{\xi} ) \right\}^2 \right] \  ,
\end{eqnarray}
to give
\begin{eqnarray}
   Z &=& \int_{-\infty}^{\infty} \frac{d y} {\sqrt{2\pi}}
   \int d^N \mbit{\xi} d^N \mbit{\xi}^*
   \exp \left[ -\frac{y^2}{2}
   -\left( \omega+\frac{\lambda}{\sqrt{N}} y \right)
   (\mbit{\xi}^* \negthickspace \cdot \mbit{\xi})
   \right]  \nonumber \\
   &\stackrel{ d^N \mbit{\xi} d^N \mbit{\xi}^*}{=}& \int_{-\infty}^{\infty} \frac{d y}{\sqrt{2\pi}}
\left( \omega+\frac{\lambda}{\sqrt{N}} y \right)^N
   \exp \left[-\frac{y^2}{2}  \right]  \nonumber \\
&\stackrel{y \to \sqrt{N} y}{=}&
   \int_{-\infty}^{\infty} \sqrt{\frac{N}{2\pi}} d y
   \exp [-N I(y)] \  , \label{base0}
\end{eqnarray}
with
\begin{eqnarray}
I(y) &\equiv& \frac{y^2}{2}-\ln (\omega+\lambda y) \ .
\end{eqnarray}
Find a solution $y_0$ of $\displaystyle{\frac{d I}{dy} = 0}$ , called
the classical solution or the saddle point, then expand $I(y)$ around $y_0$
\begin{eqnarray}
I(y) = I_0 + \frac{I^{(2)}_0}{2} (y-y_0)^2 + \frac{I^{(3)}_0}{3!} (y-y_0)^3 + \cdots \ ; \quad  I^{(n)}_0 \equiv I^{(n)}(y_0) \ ; 
\end{eqnarray}
and finally make a change of variable such that $\displaystyle{(y-y_0) \mapsto \frac{y}{\sqrt{N}} }$ to obtain
\begin{eqnarray}
Z = \exp \left( - N I_0 \right) \int_{-\infty}^\infty  \frac{dy}{\sqrt{2 \pi }}  \exp \left[ -\frac{1}{2} I^{(2)}_0 y^2 + \sqrt{\frac{1}{N}}\frac{I^{(3)}_0}{3!} y^3 + O\! \left( \frac{1}{N} \right)  \right]  \ .
\end{eqnarray}
The rest of the work is to perform a perturbation with respect to $1/N$ which is also called the loop expansion parameter. The loop expansion is well known as the semi-classical or WKB approximation. In almost all cases, $I^{(2)}_0 \neq 0$ so that $1/N$ expansion can be performed with the aid of the Gaussian integration, however there is an interesting situation: $I^{(2)}_0 =0$ , when $\lambda^2 <0$ and $|\lambda | = \omega /2$ , called caustic \cite{rf:Schulman, rf:HMT}. Utilizing a standard prescription with the Airy function in this region and $1/N$ expansion under the Gaussian integration in the other region, we can conclude that AFM does work satisfactory even within the next leading approximation but more excellently if higher order effects would be taken into account.

In this paper, we pursue the study for checking validity of AFM: models are 1-dimensional four-fermi ones with $N$ species, that is, quantum mechanical four-fermi models which can be solved analytically. We study two types of model: one has the anti-normal ordered form; since whose Hamiltonian in path integral formulas is expressed as the function of the Grassmann numbers , $\xi^*_j\xi_j$ not of $\xi^*_j\xi_{j-1}$ (the normal ordered form) or of $\xi^*_j(\xi_j + \xi_{j-1})/2 $ (the Weyl ordered form) \cite{rf:KOS4}. We establish the result that AFM works well by taking higher orders in the loop expansion as was the cases in the bosonic~\cite{rf:kashiwa} and the 0-dimensional fermi cases~\cite{rf:KS2}, which is the content of the section 2. The second model is a simpler model of the number operator. Classically there is no difference between these two models but here arises an interesting situation: all the higher order corrections seems to vanish (although we have checked it up to the two-loop order), which reminds us of the WKB exact models (see the papers \cite{rf:FFKS} and the references therein). The result that the lowest order approximation (because all the higher orders do not contribute) well fits to the exact value is also found, which is the content of the section 3. The final section is devoted to the discussion where the failure of the path integral representation in the continuous time and a new trial toward chiral symmetry breaking in QED are presented. In the appendix, the exact energy eigenvalue of the ground state is discussed.


\section{The Model--(1)}
The starting Hamiltonian is 
\begin{eqnarray}
 H(\hat{\mbit{a}}^{\dagger}, \hat{\mbit{a}})
  = -\omega \hat{\mbit{a}} \cdot \hat{\mbit{a}}^{\dagger}
  +\frac{\lambda^2}{2 N} \sum_{i,j=1}^N \hat{a}_i \hat{a}_j 
   \hat{a}_j^{\dagger} \hat{a}_i^{\dagger} \ ,
\label{Hamilton}
\end{eqnarray}
where\footnote{We employ the anti-normal ordering in the Hamiltonian (\ref{Hamilton}) because in the path integral representation we can have matched subscripts between $\mbit{\xi }_j$ and $\mbit{\xi }_j^*$: see \cite{rf:KOS4} and Eq.~(\ref{PF}).
} 
\begin{eqnarray}
\hat{\mbit{a}} \cdot \hat{\mbit{a}}^{\dagger}
    \equiv  \sum_{i=1}^N \hat{a}_i \hat{a}_i^{\dagger} \  ,
\end{eqnarray} 
$\hat{a}_i^{\dagger}$ and $\hat{a}_i$ $(i=1,2,\cdots, N)$ are the creation and the annihilation operator 
of $i$-th fermion, satisfying 
\begin{eqnarray}
   \{\hat{a}_i,\hat{a}_j^{\dagger} \} = \delta_{ij}  \ ,  \quad 
   \{\hat{a}_i,\hat{a}_j \} = \{\hat{a}_i^{\dagger} ,\hat{a}_j^{\dagger} \} = 0  \ ; 
   \quad  (i,j=1,2,\cdots, N) \ . 
\end{eqnarray}
Introduce the number operator
\begin{eqnarray}
 \hat{\mbit{n}} \equiv \hat{\mbit{a}}^{\dagger} \cdot \hat{\mbit{a}} \left( = \sum_{i=1}^N \hat{a}_i^{\dagger} \hat{a}_i \right)  \ ,
\end{eqnarray}
whose eigenstates are 
\begin{eqnarray}
& & \hat{\mbit{n}}|n, r \rangle = n |n, r \rangle  \ ;  \quad  \left( n=0, 1,2, \dots, N \  ; \  r = 1,2, \dots, \left(
                                                                      \begin{array}{c}
                                                                       	N \\ 
                                                                       	n
                                                                      \end{array}
                                                                      \right)  \right) \\ 
& &  |n, r \rangle \equiv \hat{a}_{i_1}^{\dagger} \hat{a}_{i_2}^{\dagger} \cdots \hat{a}_{i_n}^{\dagger} | 0 \rangle  \ ; \qquad \left( i_1 > i_2 > \cdots > i_n \right)  \ ; 
\end{eqnarray}
where $r$ specifies the number of combinations of $N$ elements taken $n$ at a time without repetition. 

The Hamiltonian (\ref{Hamilton}) reads in terms of $\hat{\mbit{n}}$ as
\begin{eqnarray}
 H(\hat{\mbit{a}}^{\dagger}, \hat{\mbit{a}})
  &=& -\left(\omega - \frac{\lambda^2}{2} + \frac{\lambda^2}{2 N} \right) N  
  + \left(\omega - \lambda^2 + \frac{\lambda^2}{2 N} \right) 
  \hat{\mbit{n}} 
  + \frac{\lambda^2}{2 N} \hat{\mbit{n}}^2 \ ,
\end{eqnarray}
whose eigenvalue, $H(\hat{\mbit{a}}^{\dagger},\hat{\mbit{a}}) | n ,r  \rangle  = E_n | n ,r  \rangle$ , is
\begin{eqnarray}
E_n = -\left(\omega - \frac{\lambda^2}{2} + \frac{\lambda^2}{2 N} \right) N 
  + \left(\omega - \lambda^2 + \frac{\lambda^2}{2 N} \right) n
  + \frac{\lambda^2}{2 N} n^2 \ . 
\label{Eeq}
\end{eqnarray}
The ground state energy $E_{\mbox{\scriptsize{G}}}$ is given as the lowest energy state out of $N+1$ eigenenergies; 
\begin{eqnarray}
 E_{\mbox{\scriptsize{G}}} = \min E_n \ , \qquad (n=0, 1, 2, \dots , N) \ ; \label{GroundStateEnergy}
\end{eqnarray}
whose explicit calculations are relegated to the appendix.

Meanwhile $E_{\mbox{\scriptsize{G}}}$ can be picked up from the partition function $Z(T)$,
\begin{eqnarray}
Z(T) \equiv {\rm Tr} \left( \mathrm{e}^{- T H } \right) = \sum_{n=0}^N \sum_{r=1}^{\mbox{\tiny $\left(
  \begin{array}{c}
  N \\ 
  n
 \end{array}
 \right)$}}\langle n,r |\mathrm{e}^{- T H }   |n, r \rangle  \ ;  
\end{eqnarray}
such that 
\begin{eqnarray}
 E_{\mbox{\scriptsize{G}}} = -\lim_{T \to \infty} \frac{1}{T} \ln Z(T) 
  = -\lim_{T \to \infty} \frac{1}{T} \ln 
\left(\sum_{n=0}^N 
 \left( 
  \begin{array}{c} 
   N \\
   n 
  \end{array} 
 \right) 
 \mathrm{e}^{-T E_n} \right) \ .
\label{exactgroundstateenergy}
\end{eqnarray}
whose final expression apparently coincides with the definition (\ref{GroundStateEnergy}). $Z(T)$ has a path integral representation,
\begin{eqnarray}
 Z(T) &=& \lim_{N_t \to \infty} \prod_{j=1}^{N_t} 
  \int \int d^N \mbit{\xi}_j d^N \mbit{\xi}_j^*   \nonumber \\
&&   \times  \exp \left[ -\sum_{j=1}^{N_t} 
  \left\{ (\mbit{\xi}_j^* \cdot \Delta \mbit{\xi}_j) 
  +\Delta t \left\{ \omega (\mbit{\xi}_j^* \cdot \mbit{\xi}_j) 
  +\frac{\lambda^2}{2 N}(\mbit{\xi}_j^* \cdot \mbit{\xi}_j)^2 
  \right\} \right\} \right]_{\mbox{\scriptsize{AP}}} \ , \label{PF}
\end{eqnarray} 
where $\Delta t \equiv T/ N_t$ and AP stands for the anti-periodic boundary condition, $\mbit{\xi}_0 = -\mbit{\xi}_{N_t}$ \cite{rf:KOS3}. Estimating $Z(T)$ under AFM and comparing the results with the exact value, we can check validity of AFM.

\begin{figure}[ht]
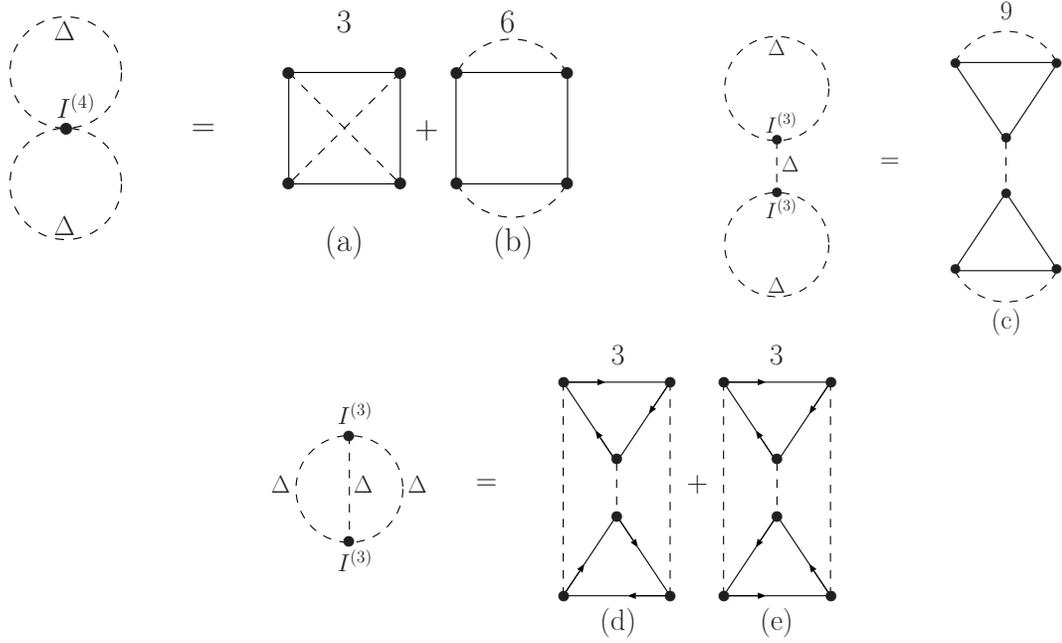

\begin{eqnarray*}
& & \raisebox{1cm}{\epsfxsize=7.5cm \epsfbox{graph2looptype1.epsi}} \hspace{20mm}\epsfxsize=4.5cm \epsfbox{graph2looptype2.epsi} \\
& & \hspace{35mm} \epsfxsize=7.5cm \epsfbox{graph2looptype3.epsi}
\end{eqnarray*} 
\caption{Two-loop graphs: the vertices, $I^{(3)}$ (\ref{I3}) and $I^{(4)}$ (\ref{I4}) are contracting to the points in the left hand side to recognize the two-loop explicitly. $\Delta$, the dotted line, denotes the propagator of the auxiliary field. In the right hand side, the line designates the fermion propagator (\ref{FerPro}) and the numbers upper in (a) $\sim$ (e) denote multiplicity.}
\label{fig:2loopver1}
\end{figure}

Introducing the auxiliary field $\sigma_j$ in terms of the Gaussian identity 
\begin{eqnarray}
 1 = \lim_{N_t \to \infty} \prod_{j=1}^{N_t} \int_{-\infty}^{\infty} 
  \sqrt{\frac{\Delta t}{2 \pi}} d \sigma_j  
  \exp \bigg[-\frac{\Delta t}{2} \sum_{j=1}^{N_t} \left\{ \sigma_j 
  + \frac{i \lambda}{\sqrt{N}}
  (\mbit{\xi}_j^* \cdot \mbit{\xi}_j ) \right\}^2 \bigg] \ ,
\end{eqnarray}
to erase the $(\mbit{\xi}^* \cdot \mbit{\xi} )^2$ term in Eq.~(\ref{PF}) and performing the Grassmann integration $d^N \mbit{\xi}_j d^N \mbit{\xi}_j^*$, we obtain
\begin{eqnarray}
 Z(T) &=& \lim_{N_t \to \infty} \prod_{j=1}^{N_t} 
  \int \sqrt{\frac{N \Delta t}{2 \pi}} d \sigma_j
  \exp[-NI[\sigma]] \ , \label{PF2}
\end{eqnarray} 
where 
\begin{eqnarray}
 I[\sigma] = \frac{\Delta t}{2} \sum_{j=1}^{N_t} \sigma_j^2
  -\ln \det \left\{ \delta_{ij}-\delta_{i-1,j}+\Delta t
  \left( \omega + i \lambda \sigma_i \right) 
  \delta_{ij} \right\} \ . \label{action}
\end{eqnarray}  
As was explained in the introduction $1/N$ is the loop expansion parameter. When $N$ goes larger (although we will assume that $N$ is not large in the following), it is expected that the integral is dominated by the saddle point $\sigma_{0j}$ obeying the equation of motion,
\begin{eqnarray}
 0 = I^{(1)}_j[\sigma_0] 
\equiv \left. \frac{\delta I[\sigma]}{\delta \sigma_j} 
\right|_{\sigma = \sigma_0}
= \Delta t ( \sigma_{0j}-i \lambda S_{jj}) \ ,
  \label{classsol}
\end{eqnarray}
where the Fermion propagator $S_{ij}$ obeys
\begin{eqnarray}
 \sum_{k=1}^{N_t} \left\{ \delta_{ik}-\delta_{i-1,k}+\Delta t
  \left( \omega + i \lambda \sigma_{0i} \right) 
\delta_{ik} \right\} S_{kj}=\delta_{ij} \ .  \label{FerPro}
\end{eqnarray}
Expand $I[\sigma]$ around $\sigma_{0j}$ such that
\begin{eqnarray}
I[\sigma] &=& I[\sigma_0]
  +\frac{1}{2!}I^{(2)}_{ij}(\sigma_i-\sigma_{0i})(\sigma_j-\sigma_{0j}) 
  \nonumber \\
 &&+\sum_{k=3}^{\infty} \frac{1}{k!}\sum_{j_1,j_2,\cdots,j_k=1}^{N_t} 
  I^{(k)}_{j_1 j_2 \cdots j_k} (\sigma_{j_1}-\sigma_{0j_1})
  (\sigma_{j_2}-\sigma_{0j_2}) \cdots (\sigma_{j_k}-\sigma_{0j_k}) \ ,  \label{ExpansionI}
\end{eqnarray}
where we have used the abbreviations,
\begin{eqnarray}
 I^{(k)}_{j_1 \cdots j_k} &\equiv& \left. \frac{\delta^k I[\sigma]}
  {\delta \sigma_{j_1} \cdots \delta \sigma_{j_k}} 
\right|_{\sigma=\sigma_0} \ .
\end{eqnarray}
Shifting and scaling the integration variables, we obtain
\begin{eqnarray}
 Z(T) &=& 
  \mathrm{e}^{-NI[\sigma_0]} \lim_{N_t \to \infty} \prod_{j=1}^{N_t} 
  \int \sqrt{\frac{ \Delta t}{2 \pi}} d \sigma_j
  \exp \bigg[-\frac{1}{2!}  \sum_{i, j=1}^{N_t}  I^{(2)}_{ij}  \sigma_i \sigma_j    \nonumber \\
 && \hspace{20mm} -  
\sum_{k=3}^{\infty} \frac{1}{k!N^{k/2-1}}\sum_{j_1,j_2,\cdots,j_k=1}^{N_t} 
  I^{(k)}_{j_1 j_2 \cdots j_k} \sigma_{j_1} \sigma_{j_2} \cdots \sigma_{j_k}
 \bigg] \ , \label{PF3}
\end{eqnarray}
where 
\begin{eqnarray}
 I^{(2)}_{ij}  \equiv  \Delta t \Delta^{-1}_{ij} 
= \Delta t ( \delta_{ij}-\lambda^2(\Delta t) S_{ij}S_{ji}) \ ,   \label{I2}  
\end{eqnarray}
with $\Delta_{ij}$ being a propagator of the auxiliary field, and
\begin{eqnarray} 
 I^{(3)}_{ijk} &=& i \lambda^3 (\Delta t)^3  (S_{ij}S_{jk}S_{ki}+S_{ik}S_{kj}S_{ji}) \ , \label{I3} \\ 
 I^{(4)}_{ijkl} &=& \lambda^4 (\Delta t)^4 
  (S_{ij}S_{jk}S_{kl}S_{li}+S_{ij}S_{jl}S_{lk}S_{ki}
  +S_{ik}S_{kj}S_{jl}S_{li} \nonumber \\
 &&  \hspace{10mm}+ S_{ik}S_{kl}S_{lj}S_{ji}
   +S_{il}S_{lj}S_{jk}S_{ki}+S_{il}S_{lk}S_{kj}S_{ji}) \ . \label{I4}
\end{eqnarray}
Armed with these we have
\begin{eqnarray}
Z_{\mbox{\scriptsize{2-loop}}} = \exp \left[-N I[\sigma_0] -\frac{1}{2}\ln \det 
\Delta^{-1}_{ij} \right] \left[1+ \frac{1}{N} \left( \mbox{two-loop graphs} \right) \right] \ , \label{Z2-loop}
\end{eqnarray} 
where the two-loop graphs are shown in Fig.~\ref{fig:2loopver1}, the tree part and the tree with the one-loop contribution are given as
\begin{eqnarray}
Z_{\mbox{\scriptsize{tree}}} &=& \mathrm{e}^{-N I[\sigma_0]}  \ ; \label{Z-tree}\\
 Z_{\mbox{\scriptsize{1-loop}}} &=& \exp \left[ -N I[\sigma_0] -\frac{1}{2}\ln \det 
\Delta^{-1}_{ij} \right] \  ; \label{Z1-loop} 
\end{eqnarray}
respectively.

\begin{figure}[h]
$$
\epsfxsize=7cm \epsfbox{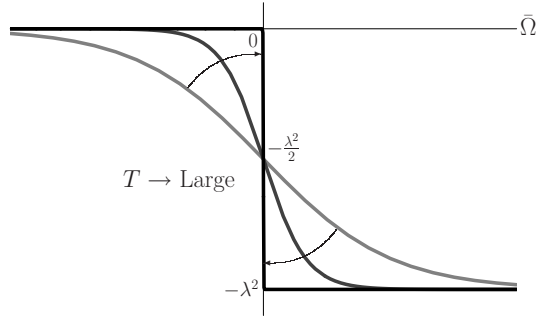} 
$$
\caption{The shapes of the right hand side of the gap equation (\ref{gapeq2}) when $\lambda^2 > 0 $. Those go to the step function when $T \mapsto \infty$.}
\label{fig:GapForm}
\end{figure}

Now let us start a detailed estimation: take a time-independent solution, 
\begin{eqnarray}
 \sigma_{0j} \mapsto \bar{\sigma}_0  \   ;      
\end{eqnarray}  
with the overlined symbol, since we are interested in the ground state (=the vacuum). The Fermion propagator (\ref{FerPro}) can be calculated with the aid of the anti-periodic eigenfunctions $f_p(0) = - f_p(N_t)$:
\begin{eqnarray}
 f_p(j) = \frac{1}{\sqrt{N_t}}  \mathrm{e}^{i \pi (2p+1) j/N_t} \ ; \qquad 
(1 \leq j \leq N_t \  , \quad 1 \leq p \leq N_t) \ , \label{antiperiodic}   
\end{eqnarray}
obeying
\begin{eqnarray}
\sum_{p=1}^{N_t} f_p(j)f_p^*(k) = \delta_{jk} \ , \qquad 
\sum_{j=1}^{N_t} f_p^*(j)f_q(j) = \delta_{pq} \ , 
\end{eqnarray}
as
\begin{eqnarray}
 \bar{S}_{jk} &=& \sum_{p=1}^{N_t} f_p^*(j)\bar{S}(p)  f_p(k) = -\frac{1}{{N_t}} \sum_{p=1}^{N_t} \frac{\left(x_p\right)^{k-j}}
  {x_p-(1+\bar{\Omega} \Delta t)} \nonumber \\
 &=& \frac{1}{(1+\bar{\Omega} \Delta t)^{N_t} +1}
  \left\{ \theta_{jk} (1+\bar{\Omega} \Delta t)^{{N_t}+k-j-1} 
   - \theta_{k,j+1} (1+\bar{\Omega} \Delta t)^{k-j-1} \right\} \ ,
\label{Greenfunction}
\end{eqnarray}
where 
\begin{eqnarray}
\bar{\Omega} \equiv  \omega + i \lambda \bar{\sigma}_0 \ , \label{Omegabar}
\end{eqnarray}
\begin{eqnarray}
\theta_{ij} = \left\{
              \begin{array}{c}
               1  \ : \qquad  i \geq  j 	\\
               \noalign{\vspace{1mm}}
               0  \ : \qquad  i <  j    
              \end{array}
              \right.    \hspace{10mm} \  ;  
\end{eqnarray}
and 
\begin{eqnarray}
 x_p \equiv \mathrm{e}^{i \pi (2p+1) /N_t}  \ , 
\end{eqnarray}
is the $N_t$th root of $(-1)$. The equation of motion (\ref{classsol}), now called the gap equation, becomes with the help of the propagator (\ref{Greenfunction})
\begin{eqnarray}
 \bar{\sigma}_0 = i \lambda \bar{S}_{ii} 
= i \lambda \frac{(1+\bar{\Omega} \Delta t)^{{N_t}-1}}
{(1+\bar{\Omega} \Delta t )^{N_t}+1}
\stackrel{{N_t} \to \infty}{\longrightarrow}
i \lambda \frac{\mathrm{e}^{\bar{\Omega} T}}{\mathrm{e}^{\bar{\Omega} T}+1} \ .
\end{eqnarray}
In view of Eq.~(\ref{Omegabar}), this reads
\begin{eqnarray}
 \bar{\Omega}-\omega
= - \lambda^2 \frac{\mathrm{e}^{\bar{\Omega} T}}{\mathrm{e}^{\bar{\Omega} T}+1} \ ,
\label{gapeq2}
\end{eqnarray}
whose right hand side becomes a step function as $T \mapsto \infty$ (See Fig.~\ref{fig:GapForm}).

The overlined propagator $\bar{\Delta}_{ij}$ of the auxiliary field
(\ref{I2}) can be expressed as 
\begin{eqnarray}
\bar{\Delta}_{ij}= \sum_{p=1}^{N_t}F_p^*(i) \bar{\Delta}(p) F_p(j)  \ ;  \quad  
\bar{\Delta}(p)  \equiv  \frac{1}{a}  -\frac{T b}{a(a+T b)} \delta_{p N_t}  \ ; \label{mesonpro}
\end{eqnarray}
where
\begin{eqnarray}
a \equiv 1- \Delta t \lambda^2 
\frac{(1+\bar{\Omega} \Delta t)^{N_t-2}}{(1+\bar{\Omega} \Delta t)^{N_t}+1} \ , \qquad 
b \equiv  \lambda^2 \frac{(1+\bar{\Omega} \Delta t)^{N_t-2}}
{\{(1+\bar{\Omega} \Delta t)^{N_t}+1 \}^2} \ ,
\end{eqnarray}
and use has been made of the periodic eigenfunctions $F_p(0) = F_p(N_t)$:
\begin{eqnarray}
 F_p(j) = \frac{1}{\sqrt{N_t}}  \mathrm{e}^{2 \pi i p j/N_t} \ ; \qquad 
(1 \leq j \leq N_t \  , \quad 1 \leq p \leq N_t) \ , \label{periodic}   
\end{eqnarray}
with
\begin{eqnarray}
\sum_{p=1}^{N_t} F_p(j)F_p^*(k) = \delta_{jk} \ , \qquad 
\sum_{j=1}^{N_t} F_p^*(j)F_q(j) = \delta_{pq} \ .
\end{eqnarray}
Therefore the tree part (\ref{Z-tree}) is found as
\begin{eqnarray}
\bar{Z}(T)_{\mbox{\scriptsize{tree}}} = \exp (-N \bar{I}[\bar{\sigma}_0])  \ ;  
\end{eqnarray} 
where
\begin{eqnarray}
\bar{I}[\bar{\sigma}_0] &=& \frac{T}{2} \bar{\sigma}_0^2 
- \ln \det \{\delta_{ij}-\delta_{i-1,j}+\Delta t \bar{\Omega} \}  \nonumber \\ 
&=&
-\frac{T}{2\lambda^2}(\bar{\Omega}-\omega)^2
-\ln (1+\mathrm{e}^{\bar{\Omega} T})  \  .
\end{eqnarray}
We need to know $\det \bar{\Delta}_{ij}^{-1}$ in order to obtain the one-loop part (\ref{Z1-loop}): 
\begin{eqnarray}
 \det \bar{\Delta}^{-1}_{ij} &=& \left[1- \Delta t \lambda^2 
\frac{(1+ \bar{\Omega} \Delta t)^{{N_t}-2}}
{(1+ \bar{\Omega} \Delta t)^{N_t} + 1} \right]^{{N_t}-1} \nonumber \\
&& \times \left[1-\Delta t \lambda^2 
\frac{(1+ \bar{\Omega} \Delta t)^{{N_t}-2}}
{(1+ \bar{\Omega} \Delta t)^{N_t} + 1}
+ T \lambda^2 \frac{(1+ \bar{\Omega} \Delta t)^{{N_t}-2}}
{\left\{(1+ \bar{\Omega} \Delta t)^{N_t} + 1 \right\}^2 } \right] \nonumber  \\
& \stackrel{{N_t} \to \infty}{\longrightarrow}&
\exp \left[- \lambda^2 \frac{\mathrm{e}^{\bar{\Omega} T}}{\mathrm{e}^{\bar{\Omega} T}+1 }
T \right] \left[1+\lambda^2 \frac{\mathrm{e}^{\bar{\Omega} T}}{\left( \mathrm{e}^{\bar{\Omega} T}+1 \right)^2 } T \right] \ .  \label{OneLoopPart}
\end{eqnarray}
Note the term proportional to $\Delta t$ in the second line vanishes when $N_t \mapsto \infty$.

The two-loop parts shown in Fig.~\ref{fig:2loopver1} can be calculated by putting the overlined quantities, $\bar{\Delta}_{ij}$ and $\bar{S}_{ij}$ into $I^{(3)}_{ijk}$, (\ref{I3}), and $I^{(4)}_{ijkl}$, (\ref{I4}): 
\begin{eqnarray}
\mbox{Fig.~\ref{fig:2loopver1}(a) } &=& -\frac{\lambda^4}{4} (\Delta t)^2 
\sum_{i,j,k,l} \bar{S}_{ij} \bar{S}_{jk} \bar{S}_{kl} \bar{S}_{li} 
\bar{\Delta}_{ik} \bar{\Delta}_{jl} \nonumber \\
&\stackrel{{N_t} \to \infty}{\longrightarrow}& 
-\frac{\lambda^4 T^2 \mathrm{e}^{2 \bar{\Omega} T}}
{4 (\mathrm{e}^{ \bar{\Omega} T}+1)^4}
+\frac{\lambda^6 T^3 \mathrm{e}^{3 \bar{\Omega} T}}
{2 (\mathrm{e}^{ \bar{\Omega} T}+1)^6} 
\left( 1+\frac{ \lambda^2 T \mathrm{e}^{ \bar{\Omega} T}}
{ (\mathrm{e}^{ \bar{\Omega} T}+1)^2} \right)^{-1} \nonumber \\
&&+\frac{\lambda^8 T^4 (\mathrm{e}^{2 \bar{\Omega} T}-4 \mathrm{e}^{ \bar{\Omega} T}+1) 
\mathrm{e}^{3 \bar{\Omega} T}}
{24 (\mathrm{e}^{ \bar{\Omega} T}+1)^8} 
\left( 1+\frac{ \lambda^2 T \mathrm{e}^{ \bar{\Omega} T}}
{ (\mathrm{e}^{ \bar{\Omega} T}+1)^2} \right)^{-2} \ . \label{TwoLoopPartA}
\end{eqnarray}
\begin{eqnarray}
\mbox{Fig.~\ref{fig:2loopver1} (b)} &=& -\frac{\lambda^4}{2} (\Delta t)^2 
 \sum_{i,j,k,l} \bar{S}_{ij} \bar{S}_{jk} \bar{S}_{kl} \bar{S}_{li} 
\bar{\Delta}_{ij} \bar{\Delta}_{kl} \nonumber \\
&\stackrel{{N_t} \to \infty}{\longrightarrow}& 
\frac{\lambda^4 T^2 \mathrm{e}^{3 \bar{\Omega} T}}
{2 (\mathrm{e}^{ \bar{\Omega} T}+1)^4}
-\frac{\lambda^6 T^3 (\mathrm{e}^{ \bar{\Omega} T}-1) \mathrm{e}^{3 \bar{\Omega} T}}
{2 (\mathrm{e}^{ \bar{\Omega} T}+1)^6} 
\left( 1+\frac{ \lambda^2 T \mathrm{e}^{ \bar{\Omega} T}}
{ (\mathrm{e}^{ \bar{\Omega} T}+1)^2} \right)^{-1} \nonumber \\
&&+\frac{\lambda^8 T^4 (\mathrm{e}^{2 \bar{\Omega} T}-4 \mathrm{e}^{ \bar{\Omega} T}+1) 
\mathrm{e}^{3 \bar{\Omega} T}}
{12 (\mathrm{e}^{ \bar{\Omega} T}+1)^8} 
\left( 1+\frac{ \lambda^2 T \mathrm{e}^{ \bar{\Omega} T}}
{ (\mathrm{e}^{ \bar{\Omega} T}+1)^2} \right)^{-2} \ .  \label{TwoLoopPartB}
\end{eqnarray}
\begin{eqnarray}
\mbox{Fig.~\ref{fig:2loopver1}(c)} &=& -\frac{\lambda^6}{2} (\Delta t)^3 
 \sum_{i,j,k,i',j',k'}  \bar{S}_{ij} \bar{S}_{jk} \bar{S}_{ki}
 \bar{S}_{i'j'} \bar{S}_{j'k'} \bar{S}_{k'i'}  
\bar{\Delta}_{ij} \bar{\Delta}_{kk'} \bar{\Delta}_{i'j'} \nonumber \\
&\stackrel{{N_t} \to \infty}{\longrightarrow}& 
-\frac{\lambda^6 T^3 \mathrm{e}^{4 \bar{\Omega} T}}
{2 (\mathrm{e}^{ \bar{\Omega} T}+1)^6}
+\frac{\lambda^8 T^4 (2 \mathrm{e}^{\bar{\Omega} T}-1)\mathrm{e}^{4 \bar{\Omega} T}}
{2 (\mathrm{e}^{ \bar{\Omega} T}+1)^8} 
\left( 1+\frac{ \lambda^2 T \mathrm{e}^{ \bar{\Omega} T}}
{ (\mathrm{e}^{ \bar{\Omega} T}+1)^2} \right)^{-1} \nonumber \\
&&-\frac{\lambda^{10} T^5 (\mathrm{e}^{ \bar{\Omega} T}-1) (5\mathrm{e}^{ \bar{\Omega} T}-1) 
\mathrm{e}^{4 \bar{\Omega} T}}
{8 (\mathrm{e}^{ \bar{\Omega} T}+1)^{10}} 
\left( 1+\frac{ \lambda^2 T \mathrm{e}^{ \bar{\Omega} T}}
{ (\mathrm{e}^{ \bar{\Omega} T}+1)^2} \right)^{-2} \nonumber \\
&  &  +\frac{\lambda^{12} T^6 (\mathrm{e}^{ \bar{\Omega} T}-1)^2  
\mathrm{e}^{5 \bar{\Omega} T}}
{8 (\mathrm{e}^{ \bar{\Omega} T}+1)^{12}} 
\left( 1+\frac{ \lambda^2 T \mathrm{e}^{ \bar{\Omega} T}}
{ (\mathrm{e}^{ \bar{\Omega} T}+1)^2} \right)^{-3} \ . \label{TwoLoopPartC}
\end{eqnarray}
\begin{eqnarray}
\mbox{Fig.~\ref{fig:2loopver1}(d)} &=& -\frac{\lambda^6}{6} (\Delta t)^3  \sum_{i,j,k,i',j',k'}
 \bar{S}_{ij} \bar{S}_{jk} \bar{S}_{ki}
 \bar{S}_{i'j'} \bar{S}_{j'k'} \bar{S}_{k'i'}  \bar{\Delta}_{ii'} \bar{\Delta}_{jk'} \bar{\Delta}_{kj'}  \nonumber \\
&\stackrel{{N_t} \to \infty}{\longrightarrow}& 
\frac{\lambda^6 T^3 \mathrm{e}^{3 \bar{\Omega} T}}
{6 (\mathrm{e}^{ \bar{\Omega} T}+1)^6}  \nonumber  \\ 
& & +\frac{\lambda^8 T^4 (\mathrm{e}^{2 \bar{\Omega} T}-4 \mathrm{e}^{ \bar{\Omega} T}+1) \mathrm{e}^{2 \bar{\Omega} T}}
{12 (\mathrm{e}^{ \bar{\Omega} T}+1)^8} 
\left( 1+\frac{ \lambda^2 T \mathrm{e}^{ \bar{\Omega} T}}
{ (\mathrm{e}^{ \bar{\Omega} T}+1)^2} \right)^{-1} \nonumber \\
&&  -\frac{\lambda^{10} T^5 (\mathrm{e}^{ \bar{\Omega} T}-1)^2 
\mathrm{e}^{4 \bar{\Omega} T}} {8 (\mathrm{e}^{ \bar{\Omega} T}+1)^{10}} 
\left( 1+\frac{ \lambda^2 T \mathrm{e}^{ \bar{\Omega} T}}
{ (\mathrm{e}^{ \bar{\Omega} T}+1)^2} \right)^{-2} \nonumber \\ 
& & -\frac{\lambda^{12} T^6 (\mathrm{e}^{ \bar{\Omega} T}-1)^2  
\mathrm{e}^{5 \bar{\Omega} T}}
{24 (\mathrm{e}^{ \bar{\Omega} T}+1)^{12}} 
\left( 1+\frac{ \lambda^2 T \mathrm{e}^{ \bar{\Omega} T}}
{ (\mathrm{e}^{ \bar{\Omega} T}+1)^2} \right)^{-3} \ . \label{TwoLoopPartD}
\end{eqnarray}
\begin{eqnarray}
\mbox{Fig.~\ref{fig:2loopver1}(e)} &=& -\frac{\lambda^6}{6} (\Delta t)^3 
  \sum_{i,j,k,i',j',k'} \bar{S}_{ij} \bar{S}_{jk} \bar{S}_{ki}
 \bar{S}_{i'j'} \bar{S}_{j'k'} \bar{S}_{k'i'}  
\bar{\Delta}_{ii'} \bar{\Delta}_{jj'} \bar{\Delta}_{kk'}  \nonumber \\
&\stackrel{{N_t} \to \infty}{\longrightarrow}& 
-\frac{\lambda^6 T^3 (\mathrm{e}^{2 \bar{\Omega} T}+1) \mathrm{e}^{2 \bar{\Omega} T}}
{12 (\mathrm{e}^{ \bar{\Omega} T}+1)^6} \nonumber  \\ 
& & +\frac{\lambda^8 T^4 (\mathrm{e}^{2 \bar{\Omega} T}-\mathrm{e}^{ \bar{\Omega} T}+1) 
\mathrm{e}^{3 \bar{\Omega} T}}
{6 (\mathrm{e}^{ \bar{\Omega} T}+1)^8} 
\left( 1+\frac{ \lambda^2 T \mathrm{e}^{ \bar{\Omega} T}}
{ (\mathrm{e}^{ \bar{\Omega} T}+1)^2} \right)^{-1} \nonumber \\
&&-\frac{\lambda^{10} T^5 (\mathrm{e}^{ \bar{\Omega} T}-1)^2 
\mathrm{e}^{4 \bar{\Omega} T}}
{8 (\mathrm{e}^{ \bar{\Omega} T}+1)^{10}} 
\left( 1+\frac{ \lambda^2 T \mathrm{e}^{ \bar{\Omega} T}}
{ (\mathrm{e}^{ \bar{\Omega} T}+1)^2} \right)^{-2} \nonumber \\ 
& & +\frac{\lambda^{12} T^6 (\mathrm{e}^{ \bar{\Omega} T}-1)^2  
\mathrm{e}^{5 \bar{\Omega} T}}
{24 (\mathrm{e}^{ \bar{\Omega} T}+1)^{12}} 
\left( 1+\frac{ \lambda^2 T \mathrm{e}^{ \bar{\Omega} T}}
{ (\mathrm{e}^{ \bar{\Omega} T}+1)^2} \right)^{-3} \ .  \label{TwoLoopPartE}
\end{eqnarray}
(Note that a difference between Fig.~\ref{fig:2loopver1} (d) and (e); the propagator $\bar{\Delta}_{jk'} \bar{\Delta}_{kj'}$ in (d) becomes $\bar{\Delta}_{jj'} \bar{\Delta}_{kk'}$ in (e).)

\vspace{2mm}

According to the definition (\ref{exactgroundstateenergy}) we obtain
\begin{eqnarray}
 E_{\mbox{\scriptsize{G}}}^{\mbox{\scriptsize{2-loop}}} &=& 
N \left[ -\frac{1}{2 \lambda^2} 
  (\bar{\Omega}-\omega)^2 - \lim_{T \to \infty} \frac{1}{T}
  \ln \left( 1 + \mathrm{e}^{\bar{\Omega} T} \right) \right] \nonumber \\
 && -\lim_{T \to \infty } 
\frac{\lambda^2}{2} \frac{\mathrm{e}^{\bar{\Omega} T}}{\mathrm{e}^{\bar{\Omega} T}+1} 
+ \lim_{T \to \infty} \frac{1}{2T} 
\ln \left(1+ \frac{\lambda^2 T \mathrm{e}^{\bar{\Omega} T}}
{(\mathrm{e}^{\bar{\Omega} T}+1)^2 } \right) \nonumber \\
&& -\frac{1}{N} \lim_{T \to \infty} \bigg[
\frac{\lambda^4 T (2 \mathrm{e}^{ \bar{\Omega} T}-1) 
\mathrm{e}^{2 \bar{\Omega} T}}
{4 (\mathrm{e}^{ \bar{\Omega}_0 T}+1)^4} 
\left( 1+\frac{ \lambda^2 T \mathrm{e}^{ \bar{\Omega} T}}
{ (\mathrm{e}^{ \bar{\Omega} T}+1)^2} \right)^{-1} \nonumber \\
&&-\frac{\lambda^6 T^2 (7 \mathrm{e}^{2 \bar{\Omega} T}-11 \mathrm{e}^{ \bar{\Omega} T}+1) 
\mathrm{e}^{2 \bar{\Omega} T}}
{12 (\mathrm{e}^{ \bar{\Omega} T}+1)^6} 
\left( 1+\frac{ \lambda^2 T \mathrm{e}^{ \bar{\Omega} T}}
{ (\mathrm{e}^{ \bar{\Omega} T}+1)^2} \right)^{-2} \nonumber \\
&&
+\frac{5 \lambda^8 T^3 (\mathrm{e}^{ \bar{\Omega} T}-1)^2  
\mathrm{e}^{3 \bar{\Omega} T}}
{24 (\mathrm{e}^{ \bar{\Omega} T}+1)^8} 
\left( 1+\frac{ \lambda^2 T \mathrm{e}^{ \bar{\Omega} T}}
{ (\mathrm{e}^{ \bar{\Omega} T}+1)^2} \right)^{-3}
\bigg]   \  ;  \label{2LoopEnergyInModel1}
\end{eqnarray}
as well as
\begin{eqnarray}
 E_{\mbox{\scriptsize{G}}}^{\mbox{\scriptsize{tree}}} & = &
N \left[ -\frac{1}{2 \lambda^2} 
  (\bar{\Omega}-\omega)^2 - \lim_{T \to \infty} \frac{1}{T}
  \ln \left( 1 + \mathrm{e}^{\bar{\Omega} T} \right) \right] \ ,  \\ 
E_{\mbox{\scriptsize{G}}}^{\mbox{\scriptsize{1-loop}}} &=& 
N \left[ -\frac{1}{2 \lambda^2} 
  (\bar{\Omega}-\omega)^2 - \lim_{T \to \infty} \frac{1}{T}
  \ln \left( 1 + \mathrm{e}^{\bar{\Omega} T} \right) \right] \nonumber \\
 && -\lim_{T \to \infty } 
\frac{\lambda^2}{2} \frac{\mathrm{e}^{\bar{\Omega} T}}{\mathrm{e}^{\bar{\Omega} T}+1} 
+ \lim_{T \to \infty} \frac{1}{2T} 
\ln \left(1+ \frac{\lambda^2 T \mathrm{e}^{\bar{\Omega} T}}
{(\mathrm{e}^{\bar{\Omega} T}+1)^2 } \right) \ . 
\end{eqnarray}

\vspace{2mm}

Estimation is made for two cases that (i) $\lambda^2 > 0$ and (ii) $\lambda^2 < 0$.

\vspace{1mm}

\begin{figure}[ht]

$$ \epsfxsize=7cm  \epsfbox{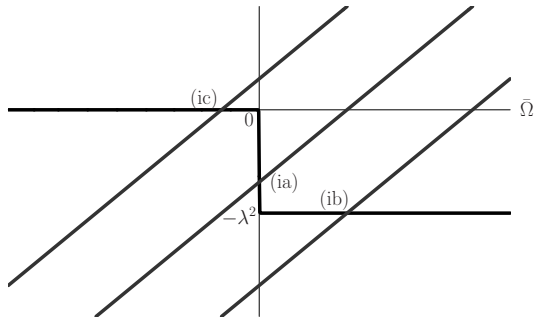}$$ 

\vspace{-2mm}

\caption{The gap equation: $\lambda^2 > 0 $ : the step function stands
 for the right hand side of the gap equation (\ref{gapeq2}) and the
 line, $\bar{\Omega } - \omega$ , for the left hand side. There is one
 crossing point in each case: (ia) ; $\omega > 0$ and 
$\lambda^2 \geq \omega $ . (ib) ; $\omega > 0$ and  
$ 0 <  \lambda^2 <  \omega $ . (ic) ; $\omega <0$ . } \label{fig:mode1Gap+}
\end{figure}

\noindent (i) $\lambda^2 > 0$ : the solution of the gap equation (\ref{gapeq2}) can be categorized into three cases, (ia), (ib), and (ic), according to the value of $\omega$: see Fig.~\ref{fig:mode1Gap+}.
\begin{itemize}

\item[(ia)] $\omega > 0$ and $\lambda^2 \geq \omega $ : $\bar{\Omega} = 0$ then from Eq.~(\ref{gapeq2}), 
$$
\mathrm{e}^{\bar{\Omega} T} = \frac{\omega}{\lambda^2-\omega} \ ,
$$ 
so that from Eq.~(\ref{2LoopEnergyInModel1})
\begin{eqnarray}
E_{\mbox{\scriptsize{G}}}^{\mbox{\scriptsize{2-loop}}} = -N \frac{\omega^2}{2 \lambda^2} 
-\frac{\omega}{2} -\frac{1}{N} \frac{\lambda^2}{8} \ , 
\end{eqnarray}
whose $O(N)$, $O(1)$, and $O(1/N)$ terms designate the tree, one-loop, and two-loop contributions, respectively.

\item[(ib)] $\omega > 0$ and $ 0 <  \lambda^2 <  \omega $ : $ \bar{\Omega} = \omega - \lambda^2 $ , then 
$$ 
\mathrm{e}^{\bar{\Omega} T} \stackrel{T \to \infty}{\longrightarrow} \infty \ ,
$$
so that there is no correction from the two-loop in Eq.~(\ref{2LoopEnergyInModel1}) to give
\begin{eqnarray} 
E_{\mbox{\scriptsize{G}}}^{\mbox{\scriptsize{1-loop}}} = 
E_{\mbox{\scriptsize{G}}}^{\mbox{\scriptsize{2-loop}}} 
= -N \left( \omega -\frac{\lambda^2}{2} \right)-\frac{\lambda^2}{2} \ , 
\end{eqnarray}
whose $O(N)$ and $O(1)$ terms are the tree and the one-loop contributions. There are two domains in $\lambda^2$ divided by $\omega$, which exposes a striking difference to the exact case (\ref{lambda+omega+Domain}) where $N+1$ domains exist. Due to these, the curve of exact energy has many discontinuities arising from the boundaries of domains. We plot the results of (ia) and (ib) with $N=2, \omega = 1 $ in the left of Fig.~\ref{fig:model1omega+lambda+}. Although the deviations in the tree and one-loop results from the exact energy is substantial, the two-loop contribution cures the situation; whose effect is much clearly seen with $N=4$ in the right of Fig.~\ref{fig:model1omega+lambda+}.
\begin{figure}[ht]
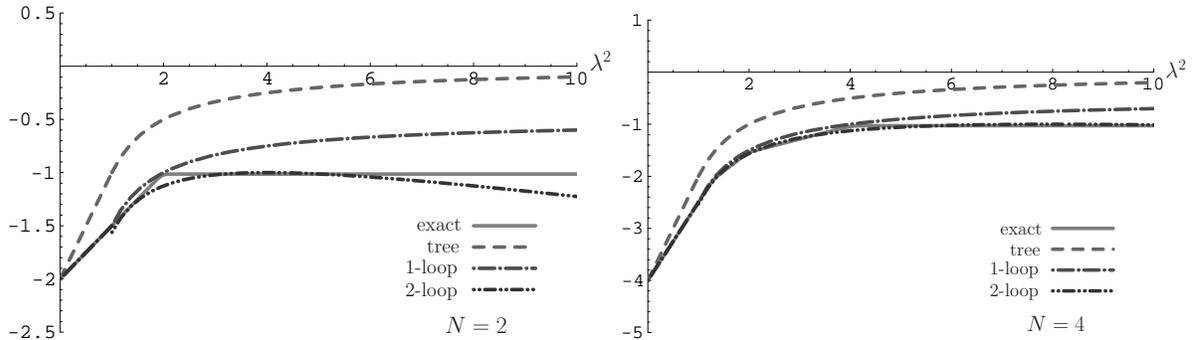

 $$
\epsfxsize=8cm  \epsfbox{model1omega+lambda+.epsi} \ \epsfxsize=7.5cm  \epsfbox{model1omega+lambda+N4.epsi} 
$$
\caption{The case of $\lambda^2 > 0 \ ; \ \omega =1$, corresponding to
 (ia) and (ib). The solid, dotted, dashed-dotted, and dashed-double dotted lines designate the exact, tree, one-loop, and two-loop results, respectively. We put $N=2(\mbox{left}), 4(\mbox{right})$. The approximation becomes better when taking higher loops into account.}
\label{fig:model1omega+lambda+}
\end{figure}

\item[(ic)] $\omega <0 : \ \bar{\Omega} = -|\omega|  $ , then
$$
\mathrm{e}^{\bar{\Omega} T} \stackrel{T \to \infty}{\longrightarrow} 0 \ ,
$$
so that the one- and two-loop results vanish, yielding to
\begin{eqnarray}   
E_{\mbox{\scriptsize{G}}}^{\mbox{\scriptsize{tree}}} = E_{\mbox{\scriptsize{G}}}^{\mbox{\scriptsize{1-loop}}} 
= E_{\mbox{\scriptsize{G}}}^{\mbox{\scriptsize{2-loop}}} = 0 \ , 
\end{eqnarray}
whose result coincides with the exact energy (\ref{ExactSOLlambda+omega-}) in the appendix.
\end{itemize}

\begin{figure}[ht]
  \begin{center}
   \raisebox{0cm}{}
   \epsfxsize=7cm  \epsfbox{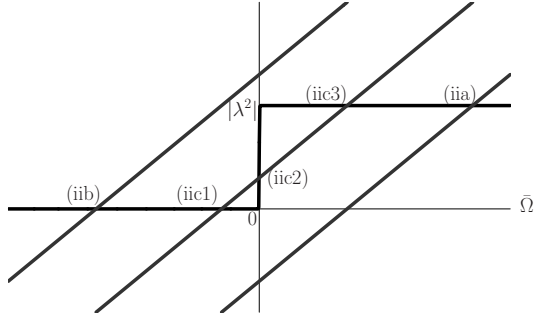} 
\end{center}
\caption{The gap equation, $\lambda^2 < 0 $ : (iia) ; $\omega > 0 $
 . (iib) ; $\omega <0$ and $0 < |\lambda^2| < |\omega| $ . (iic) ; $\omega <0$ and $|\lambda^2| \geq |\omega|  $ , where three solutions, (iic1) $\sim$ (iic3) , exist.}
\label{fig:mode1Gap-}
\end{figure}

\vspace{5mm}

\noindent (ii) $\lambda^2 < 0$ : five different solutions of Eq.~(\ref{gapeq2}) are found according the value of $\omega$. See Fig.~\ref{fig:mode1Gap-}.

\begin{figure}[ht]
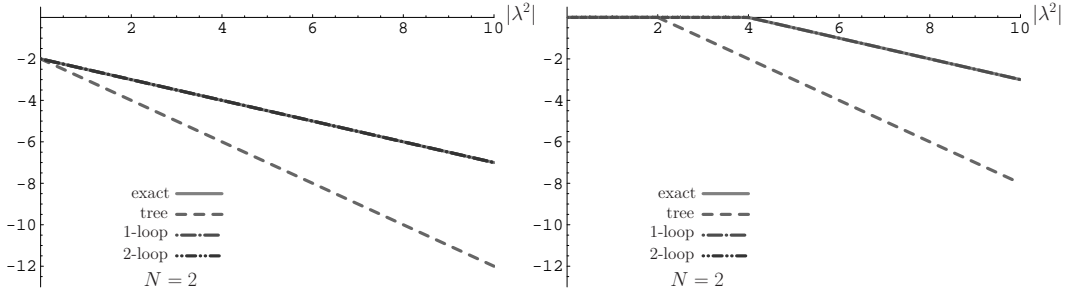

\begin{center}
\raisebox{0cm}{}
\epsfxsize=7cm \epsfbox{model1omega+lambda-.epsi} \epsfxsize=7cm \epsfbox{model1omega-lambda-.epsi}
\end{center}
\caption{Left: $\lambda^2 < 0$ ; $\omega = 1$ case, corresponding to (iia). Here also $N=2$. The solid, dotted, dashed-dotted, dashed-double dotted lines designate the exact, tree, one-loop, and two-loop results, respectively. The tree approximation is slightly deviating from the exact value but including the higher orders reproduces it: the one-loop approximation is accurate enough. \hfill Right: $\lambda^2 < 0$ ; $\omega = -1$ , and $N=2$ case, corresponding to (iib) and (iic): the two- as well as the one-loop approximation matches with the exact value.
}
\label{fig:model1omega+lambda-}
\end{figure}

\begin{itemize}

\item[(iia)] $\omega > 0 $ : $\bar{\Omega} = \omega + |\lambda^2|$ , then
$$
\mathrm{e}^{\bar{\Omega} T} \stackrel{T \to \infty}{\longrightarrow} \infty \ ,
$$
so that there is also no correction from the two-loop, yielding to
\begin{eqnarray} 
E_{\mbox{\scriptsize{G}}}^{\mbox{\scriptsize{1-loop}}} = 
E_{\mbox{\scriptsize{G}}}^{\mbox{\scriptsize{2-loop}}} 
= -N \left(  \omega + \frac{| \lambda^2|}{2}\right)  + \frac{| \lambda^2|}{2}  \ , 
\end{eqnarray}
which is nothing but the result obtained in Eq.~(\ref{ExactSOLlambda-omega+}) in the appendix.  The tree result has a slight deviation but the one-loop correction fits in the curve. (See the left graph in Fig.~\ref{fig:model1omega+lambda-}.)

\item[(iib)] $\omega <0$ and $ 0 < |\lambda^2| < |\omega|$ : $\bar{\Omega} = -|\omega | $ , then
$$
\mathrm{e}^{\bar{\Omega} T} \stackrel{T \to \infty}{\longrightarrow} 0 \ ,
$$
so that in Eq.~(\ref{2LoopEnergyInModel1}) the one- and two-loop corrections vanish to give
\begin{eqnarray} 
E_{\mbox{\scriptsize{G}}}^{\mbox{\scriptsize{tree}}} = E_{\mbox{\scriptsize{G}}}^{\mbox{\scriptsize{1-loop}}} 
= E_{\mbox{\scriptsize{G}}}^{\mbox{\scriptsize{2-loop}}} = 0 \ . 
\end{eqnarray} 

\item[(iic)] $\omega <0$ and $ |\lambda^2| \geq  |\omega|  $ : from Fig.~\ref{fig:mode1Gap-}, there are three different solutions.
\begin{itemize}
 \item[(iic1)] $\bar{\Omega} = -|\omega|$, then
$$
\mathrm{e}^{\bar{\Omega} T} \stackrel{T \to \infty}{\longrightarrow} 0 \ ,
$$
so that there is again no correction from the higher orders to obtain
\begin{eqnarray}
E_{\mbox{\scriptsize{G}}}^{\mbox{\scriptsize{tree}}} = E_{\mbox{\scriptsize{G}}}^{\mbox{\scriptsize{1-loop}}} =
 E_{\mbox{\scriptsize{G}}}^{\mbox{\scriptsize{2-loop}}} = 0 \ .
\end{eqnarray}

 \item[(iic2)] $\bar{\Omega} = 0$ , then from Eq.~(\ref{gapeq2}) 
$$ 
\mathrm{e}^{\bar{\Omega} T}=  \frac{|\omega|}{|\lambda^2|-|\omega|} \ , 
$$ 
so that
\begin{eqnarray}
 E_{\mbox{\scriptsize{G}}}^{\mbox{\scriptsize{2-loop}}} 
= N \frac{|\omega|^2}{2 |\lambda^2|} +\frac{|\omega|}{2} 
+\frac{1}{N} \frac{|\lambda^2|}{8} \ , 
\end{eqnarray} 
from Eq.~(\ref{2LoopEnergyInModel1}), which is apparently positive to be greater than (iic1) and should be discarded.

 \item[(iic3)] $ \bar{\Omega} =-|\omega| + |\lambda^2| (> 0)$ , then 
$$
\mathrm{e}^{\bar{\Omega} T} \stackrel{T \to \infty}{\longrightarrow} \infty  \ ;
$$
so that in Eq.~(\ref{2LoopEnergyInModel1}) the two-loop correction vanishes, yielding to
\begin{eqnarray}
E_{\mbox{\scriptsize{G}}}^{\mbox{\scriptsize{1-loop}}} = E_{\mbox{\scriptsize{G}}}^{\mbox{\scriptsize{2-loop}}}  
& = &  N  \left( |\omega | - \frac{| \lambda^2 |}{2} \right)  +  \frac{ | \lambda^2 |}{2}   
 \nonumber \\ 
& = &  \left\{
\begin{array}{cc}
 \mbox{positive}:	&  \displaystyle{0 < |\lambda^2| \leq \frac{2N}{N-1} |\omega| }	\\
 \noalign{\vspace{2mm}}
 \mbox{negative}:	&  \displaystyle{|\lambda^2| > \frac{2N}{N-1} |\omega| }
   \end{array}
                                                                                                                                                                                           \right.\ . 
\end{eqnarray}
\end{itemize}

\end{itemize}
Therefore we should adopt the solution (iic1) up to
$\displaystyle{|\lambda^2| = \frac{2N}{N-1} |\omega|}$ then switch it to
(iic3), whose result is plotted in the right graph of
Fig.~\ref{fig:model1omega+lambda-} with putting $N=2$, $\omega =- 1$. After including the one-loop effect, the ground state energy as well as the number of domains coincides with the exact energy~(\ref{ExactSOLlambda-omega-}) in the appendix.

\begin{figure}[ht]
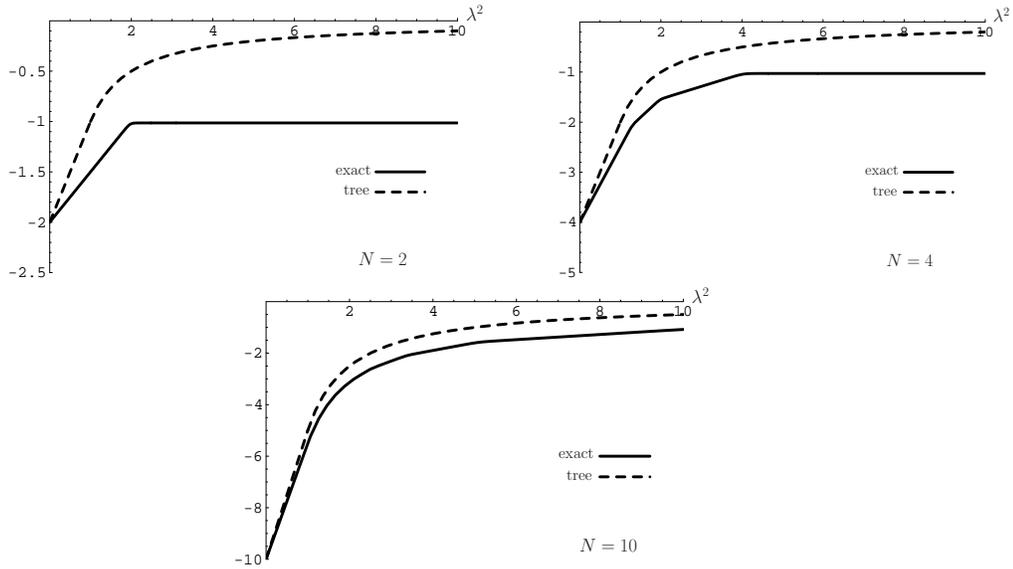


\begin{eqnarray*}
&& \epsfxsize=6.3cm \epsfbox{N2Tree.epsi}  \hspace{10mm} \epsfxsize=6cm \epsfbox{N4Tree.epsi}  \\ 
&&\hspace{30mm} \epsfxsize=6.3cm \epsfbox{N10Tree.epsi}
\end{eqnarray*}  

\caption{The exact (the solid line) and tree (the dotted line) results
 for $N=2$, $4$ and $10$ , by putting $\lambda^2 >0$ ; $\omega =1$. Discontinuities in the exact energy curve wear off gradually. Although the vertical scale (the energy value) is different, the tree results come closer and closer to the exact value.
}
\label{fig:N2 5 10Tree}
\end{figure}

\vspace{5mm}
So far the tree or one-loop result matches with the exact curve except
the case $\lambda^2 >0$ ; $\omega =1$ (ia) and (ib), where the number of
domains (two) are different from the exact one ($N+1$),
Eq.~(\ref{lambda+omega+Domain}). The two-loop effect, however, cures the
situation and the deviation becomes smaller in $N=4$
(Fig.~\ref{fig:model1omega+lambda+}). The reason is seen from
Fig.~\ref{fig:N2 5 10Tree} where we can recognize that the tree
approximation approaches closer to the exact value when $N$ moves from
$2$ to $4$ and $10$, and moreover that when $N$ goes larger,
discontinuities in the exact curve fade away. (As the number of domains
in $\lambda^2$ (\ref{lambda+omega+Domain}) increases sharpness at the
boundary wears off gradually.) Needless to say that AFM is analytic so
that unless a boundary of domains in the loop expansion happens to
coincide as of $\lambda^2 < 0$ ; $\omega < 0$, (iib) and (iic)
(Fig.~\ref{fig:model1omega+lambda-}(right)), the deviation is inevitable
as of $\lambda^2 > 0$ ; $\omega > 0$, (ia) and  (ib). Therefore we
conclude that AFM works still well even in $\lambda^2 > 0$ ; $\omega > 0$ : the deviation emerges not from the failure of AFM but from the model with discontinuities in the energy curve.

\section{The Model--(2)}
In this section, we adopt a slightly different model:
\begin{eqnarray}
 H(\hat{\mbit{a}}^{\dagger}, \hat{\mbit{a}})
  = \omega ( \hat{\mbit{a}}^{\dagger} \cdot \hat{\mbit{a}} )
  +\frac{\lambda^2}{2 N} (\hat{\mbit{a}}^{\dagger} \cdot
  \hat{\mbit{a}})(\hat{\mbit{a}}^{\dagger} \cdot \hat{\mbit{a}}) = \omega \hat{\mbit{n}} +\frac{\lambda^2}{2 N} \hat{\mbit{n}}^2  \ . 
\label{Hamiltonv2}
\end{eqnarray}
Classically, there is no difference between the model-(1) and -(2), but the reason for considering this model is that all the higher order corrections seem to vanish (although we have checked up to the two-loop) to give us another example of WKB exact model~\cite{rf:FFKS}. The energy eigenvalue is obtained as before, 
\begin{eqnarray}
& & H(\hat{\mbit{a}}^{\dagger},\hat{\mbit{a}}) | n , r \rangle = E_n | n , r \rangle  \ ;  \nonumber   \\ 
&& \hspace{5mm} E_n = \omega n + \frac{\lambda^2}{2 N} n^2 \ .   \label{Eeqv2}
\end{eqnarray}
The minimum (= the ground state) energy (\ref{GroundStateEnergy}), calculated in the appendix, should be compared with that obtained from AFM: the partition function is 
\begin{eqnarray}
 Z(T) &=& \lim_{N_t \to \infty} \prod_{j=1}^{N_t} 
  \int \int d^N \xi_j d^N \xi_j^* \exp \bigg[ -\sum_{j=1}^{N_t} 
  \bigg\{(\mbit{\xi}_j^* \cdot \Delta \mbit{\xi}_j)  \nonumber \\
  &&   \hspace{-5mm}+\Delta t 
  \left\{ N \left(\omega + \frac{\lambda^2}{2} \right) 
   + \left( \omega + \lambda^2 - \frac{\lambda^2}{2N} \right) 
   (\mbit{\xi}_j^* \cdot \mbit{\xi}_j) 
  +\frac{\lambda^2}{2 N}(\mbit{\xi}_j^* \cdot \mbit{\xi}_j)^2 
  \right\} \bigg\} \bigg]_{\mbox{\scriptsize{AP}}} \ , \label{PFv2}
\end{eqnarray} 
which, by use of the auxiliary field, turns into a new form ,
\begin{eqnarray}
 Z(T) &=& \lim_{N_t \to \infty} \prod_{j=1}^{N_t} 
  \int \sqrt{\frac{N \Delta t}{2 \pi}} d \sigma_j
  \exp[-N \tilde{I}[\sigma]] \ , \label{PF2v2}
\end{eqnarray} 
where 
\begin{eqnarray}
  \tilde{I}[\sigma] & = & T \left(\omega + \frac{\lambda^2}{2} \right)   \nonumber \\
 & &  +\frac{\Delta t}{2} \sum_{j=1}^{N_t} \sigma_j^2
  -\ln \det \left\{ \delta_{ij}-\delta_{i-1,j}+\Delta t
  \left( \omega + \lambda^2 + i \lambda \sigma_i 
   -\frac{\lambda^2}{2 N} \right) 
  \delta_{ij} \right\} \ . \label{actionv2}
\end{eqnarray}  
The saddle point $\tilde{\sigma}_{0j}$ fulfills the equation of motion
\begin{eqnarray}
 0 =  \left. \frac{\delta \tilde{I}[\sigma]}{\delta \sigma_j} 
\right|_{\sigma = \tilde{\sigma}_0}
= \Delta t (\tilde{\sigma}_{0j}-i \lambda \tilde{S}_{jj}) \ ,
  \label{classsolv2}
\end{eqnarray}
where $\tilde{S}_{jk}$ obeys
\begin{eqnarray}
 \sum_{k=1}^{N_t} \left\{ \delta_{ik}-\delta_{i-1,k}+\Delta t
  \left( \omega + \lambda^2 + i \lambda \tilde{\sigma}_{0i} 
-\frac{\lambda^2}{2 N} \right) \delta_{ik} \right\} \tilde{S}_{kj}=\delta_{ij} \ .
\label{FerProv2}
\end{eqnarray}
Note that contrary to the previous situation, $O(1/N)$ term has already appeared in the expression. Therefore the loop expansion should be distinguished from the $1/N$ expansion in this model: we should perform the loop expansion then arrange the results in the order of $1/N$. All the procedures to the $O(1/N)$ (two-loop) ground state energy are, however, the same except that all quantities are tilded; the time-independent solution of the gap equation (\ref{gapeq2}) should be expressed such that $\tilde{\sigma}_{0i} \mapsto \bar{\tilde{\sigma}}_{0}$ .

Therefore the Fermion propagator, the solution to Eq.~(\ref{FerProv2}), is given as
\begin{eqnarray}
 \bar{\tilde{S}}_{jk} = \frac{1}{(1+\bar{\tilde{\Omega}} \Delta t)^{N_t} +1}
  \left\{ \theta_{jk} (1+\bar{\tilde{\Omega}} \Delta t)^{{N_t}+k-j-1} 
   - \theta_{k,j+1} (1+\bar{\tilde{\Omega}} \Delta t)^{k-j-1} \right\} \ ,
\label{Greenfunctionv2}
\end{eqnarray}
where 
\begin{eqnarray}
\bar{\tilde{\Omega}} \equiv  \omega + \lambda^2 + i \lambda \bar{\tilde{\sigma}}_0 -\frac{\lambda^2}{2 N} \ . \label{OmegaTildeBar}
\end{eqnarray}
With the aid of the propagator (\ref{Greenfunctionv2}), the equation of motion  (\ref{classsolv2}) yields to the gap equation,
\begin{eqnarray}
 \bar{\tilde{\sigma}}_0 = i \lambda \bar{\tilde{S}}_{ii} 
= i \lambda \frac{(1+\bar{\tilde{\Omega}} \Delta t)^{{N_t}-1}}
{(1+\bar{\tilde{\Omega}} \Delta t )^{N_t}+1}
\stackrel{{N_t} \to \infty}{\longrightarrow}
i \lambda \frac{\mathrm{e}^{\bar{\tilde{\Omega}} T}}{\mathrm{e}^{\bar{\tilde{\Omega}} T}+1} \ .
\end{eqnarray}
Namely,
\begin{eqnarray}
 \bar{\tilde{\Omega}}-\left( \omega + \lambda^2 -\frac{\lambda^2}{2N} \right)
= - \lambda^2 \frac{\mathrm{e}^{\bar{\tilde{\Omega}} T}}{\mathrm{e}^{\bar{\tilde{\Omega}} T}+1} \ ;
\label{gapeq2v2}
\end{eqnarray}
in view of Eq.~(\ref{OmegaTildeBar}). In terms of $1/N$ expansion, $\bar{\tilde{\Omega}}$ can be expressed as 
\begin{eqnarray}
\bar{\tilde{\Omega}} = \bar{\Omega}_0 + \frac{\bar{\Omega}_1}{N} 
+ \frac{\bar{\Omega}_2}{N^2} + \cdots \ . \label{Omegaexpandv2}
\end{eqnarray}
Inserting this into the gap equation (\ref{gapeq2v2}) and arranging the power of $1/N$, we obtain
\begin{eqnarray}
 \bar{\Omega}_0-\left( \omega + \lambda^2 \right)
= - \lambda^2 \frac{\mathrm{e}^{\bar{\Omega}_0 T}}{\mathrm{e}^{\bar{\Omega}_0 T}+1} \ , \label{OmegazeroGap}
\end{eqnarray}
and
\begin{eqnarray}
 \bar{\Omega}_1 &=& -\frac{\lambda^2}{2} 
\left(1 + T \lambda^2 \frac{\mathrm{e}^{\bar{\Omega}_0 T}}{(\mathrm{e}^{\bar{\Omega}_0 T}+1)^2} 
\right)^{-1} \ , \\
 \bar{\Omega}_2 &=& - T^2 \left( \frac{\lambda^2}{2} \right)^3
\frac{\mathrm{e}^{\bar{\Omega}_0 T}(1-\mathrm{e}^{\bar{\Omega}_0 T}) }
{(\mathrm{e}^{\bar{\Omega}_0 T}+1)^3} 
\left(1 + T \lambda^2 \frac{\mathrm{e}^{\bar{\Omega}_0 T}}{(\mathrm{e}^{\bar{\Omega}_0 T}+1)^2} 
\right)^{-3} \ .
\end{eqnarray}

The tree part of the partition function (\ref{Z-tree}) is
\begin{eqnarray}
\bar{Z}(T)_{\mbox{\scriptsize{tree}}} = \exp (-N \bar{\tilde{I}}[\bar{\tilde{\sigma}}_0]) \ ; 
\end{eqnarray} 
with
\begin{eqnarray}
\bar{\tilde{I}}[\bar{\tilde{\sigma}}_0] &=& T\left( \omega + \frac{\lambda^2}{2} \right)
+ \frac{T}{2} \bar{\tilde{\sigma}}_0^2 
- \ln \det \{\delta_{ij}-\delta_{i-1,j}+\Delta t \bar{\tilde{\Omega}} \} \nonumber \\ 
&=& T \left( \omega + \frac{\lambda^2}{2} \right)
-\frac{T}{2\lambda^2} \left( \bar{\tilde{\Omega}} - \omega - \lambda^2 + \frac{\lambda^2}{2 N} \right) ^2
-\ln (1+\mathrm{e}^{\bar{\tilde{\Omega}} T}) \  .  \label{TreeActionInModel2}
\end{eqnarray}
The one-loop partition function (\ref{Z1-loop}) with Eq.~(\ref{OneLoopPart}) and the two-loop one shown in Figs.~\ref{fig:2loopver1} (a)-(e) or Eqs.~(\ref{TwoLoopPartA}) $\sim $ (\ref{TwoLoopPartE}), are all the same with the substitution $\bar{\Omega} \mapsto \bar{\tilde{\Omega }}$.

The ground state energy up to the two-loop is obtained from the second to the fourth line in Eq.~(\ref{2LoopEnergyInModel1}) as well as from Eq.~(\ref{TreeActionInModel2}) such that
\begin{eqnarray}
E_{\mbox{\scriptsize{G}}}^{\mbox{\scriptsize{2-loop}}} &=& N \left[    \left( \omega + \frac{\lambda^2}{2} \right)
-\frac{1}{2\lambda^2} \left( \bar{\tilde{\Omega}} - \omega - \lambda^2 + \frac{\lambda^2}{2 N} \right) ^2
- \lim_{T \to \infty }\frac{1}{T} \ln (1+\mathrm{e}^{\bar{\tilde{\Omega}} T}) \right]  \nonumber \\  
 && -\lim_{T \to \infty } 
\frac{\lambda^2}{2} \frac{\mathrm{e}^{\bar{\tilde{\Omega}} T}}{\mathrm{e}^{\bar{\tilde{\Omega}} T}+1} 
+ \lim_{T \to \infty} \frac{1}{2T} 
\ln \left(1+ \frac{\lambda^2 T \mathrm{e}^{\bar{\tilde{\Omega}} T}}
{(\mathrm{e}^{\bar{\tilde{\Omega}} T}+1)^2 } \right) \nonumber \\
&& -\frac{1}{N} \lim_{T \to \infty} \bigg[
\frac{\lambda^4 T (2 \mathrm{e}^{ \bar{\tilde{\Omega}} T}-1) 
\mathrm{e}^{2 \bar{\tilde{\Omega}} T}}
{4 (\mathrm{e}^{ \bar{\tilde{\Omega}}_0 T}+1)^4} 
\left( 1+\frac{ \lambda^2 T \mathrm{e}^{ \bar{\tilde{\Omega}} T}}
{ (\mathrm{e}^{ \bar{\tilde{\Omega}} T}+1)^2} \right)^{-1} \nonumber \\
&&-\frac{\lambda^6 T^2 (7 \mathrm{e}^{2 \bar{\tilde{\Omega}} T}-11 \mathrm{e}^{ \bar{\tilde{\Omega}} T}+1) 
\mathrm{e}^{2 \bar{\tilde{\Omega}} T}}
{12 (\mathrm{e}^{ \bar{\tilde{\Omega}} T}+1)^6} 
\left( 1+\frac{ \lambda^2 T \mathrm{e}^{ \bar{\tilde{\Omega}} T}}
{ (\mathrm{e}^{ \bar{\tilde{\Omega}} T}+1)^2} \right)^{-2} \nonumber \\
&&
+\frac{5 \lambda^8 T^3 (\mathrm{e}^{ \bar{\tilde{\Omega}} T}-1)^2  
\mathrm{e}^{3 \bar{\tilde{\Omega}} T}}
{24 (\mathrm{e}^{ \bar{\tilde{\Omega}} T}+1)^8} 
\left( 1+\frac{ \lambda^2 T \mathrm{e}^{ \bar{\tilde{\Omega}} T}}
{ (\mathrm{e}^{ \bar{\tilde{\Omega}} T}+1)^2} \right)^{-3}
\bigg]  \ .  \label{2LoopEnergyInModel2}
\end{eqnarray}
Expand $\bar{\tilde{\Omega}}$ in terms of $1/N$ according to Eq.~(\ref{Omegaexpandv2}) to give
\begin{eqnarray}
 E_{\mbox{\scriptsize{G}}}^{1/N} &=& N \left[ \omega+\frac{\lambda^2}{2}-\frac{1}{2 \lambda^2} 
  (\bar{\Omega}_0-(\omega+\lambda^2))^2 - \lim_{T \to \infty} \frac{1}{T}
  \ln \left( 1 + \mathrm{e}^{\bar{\Omega}_0 T} \right) \right] \nonumber \\
 && + \lim_{T \to \infty} \frac{1}{2T} 
\ln \left(1+ \frac{\lambda^2 T \mathrm{e}^{\bar{\Omega}_0 T}}
{(\mathrm{e}^{\bar{\Omega}_0 T}+1)^2 } \right) \nonumber \\
&& -\frac{1}{N} \lim_{T \to \infty} \bigg[
\frac{\lambda^4 T (\mathrm{e}^{2 \bar{\Omega}_0 T}-4 \mathrm{e}^{ \bar{\Omega}_0 T}+1) 
\mathrm{e}^{ \bar{\Omega}_0 T}}
{8 (\mathrm{e}^{ \bar{\Omega}_0 T}+1)^4} 
\left( 1+\frac{ \lambda^2 T \mathrm{e}^{ \bar{\Omega}_0 T}}
{ (\mathrm{e}^{ \bar{\Omega}_0 T}+1)^2} \right)^{-1} \nonumber \\
&&-\frac{\lambda^6 T^2 (4 \mathrm{e}^{2 \bar{\Omega}_0 T}-11 \mathrm{e}^{ \bar{\Omega}_0 T}+4) 
\mathrm{e}^{2 \bar{\Omega}_0 T}}
{12 (\mathrm{e}^{ \bar{\Omega}_0 T}+1)^6} 
\left( 1+\frac{ \lambda^2 T \mathrm{e}^{ \bar{\Omega}_0 T}}
{ (\mathrm{e}^{ \bar{\Omega}_0 T}+1)^2} \right)^{-2} \nonumber \\
&&
+\frac{5 \lambda^8 T^3 (\mathrm{e}^{ \bar{\Omega}_0 T}-1)^2  
\mathrm{e}^{3 \bar{\Omega}_0 T}}
{24 (\mathrm{e}^{ \bar{\Omega}_0 T}+1)^8} 
\left( 1+\frac{ \lambda^2 T \mathrm{e}^{ \bar{\Omega}_0 T}}
{ (\mathrm{e}^{ \bar{\Omega}_0 T}+1)^2} \right)^{-3}
\bigg]  \ ; \label{m2GroundStateFull}
\end{eqnarray}
where the first line is the leading approximation,
\begin{eqnarray}
 E_{\mbox{\scriptsize{G}}}^{N} = 
N \left[ \omega+\frac{\lambda^2}{2}-\frac{1}{2 \lambda^2} 
  (\bar{\Omega}_0-(\omega+\lambda^2))^2 - \lim_{T \to \infty} \frac{1}{T}
  \ln \left( 1 + \mathrm{e}^{\bar{\Omega}_0 T} \right) \right] \ , 
\end{eqnarray}
and the term including $O(1)$ approximation is
\begin{eqnarray}
 E_{\mbox{\scriptsize{G}}}^{1} &=& 
N \left[ \omega+\frac{\lambda^2}{2}-\frac{1}{2 \lambda^2} 
  (\bar{\Omega}_0-(\omega+\lambda^2))^2 - \lim_{T \to \infty} \frac{1}{T}
  \ln \left( 1 + \mathrm{e}^{\bar{\Omega}_0 T} \right) \right] \nonumber \\
 && + \lim_{T \to \infty} \frac{1}{2T} 
\ln \left(1+ \frac{\lambda^2 T \mathrm{e}^{\bar{\Omega}_0 T}}
{(\mathrm{e}^{\bar{\Omega}_0 T}+1)^2 } \right) \ . 
\end{eqnarray}
\begin{figure}[ht]
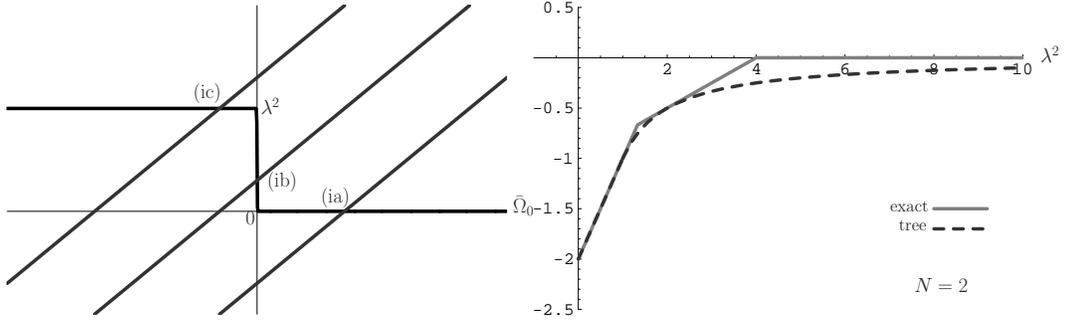

  \begin{center}
   \raisebox{0cm}{}
   \epsfxsize=7cm  \epsfbox{model2Gap+.epsi} \epsfxsize=7cm \epsfbox{model2omega-lambda+.epsi}
  \end{center}
\caption{Left: the gap equation ; $\lambda^2 > 0 $: there is one
 crossing point in each case: (ia) ; $\omega>0$. (ib) ; $\omega <0$ and
 $\lambda^2 \geq |\omega|$. (ic) ; $\omega <0$ and $0 < \lambda^2 <
 |\omega|$. \hfill Right: the ground state energy for $N=2$: 
$\lambda^2 > 0$ ; $\omega = -1$, corresponding to (ib) and (ic). All the corrections vanish and the lowest result (the dotted line) almost reproduces the exact value (the solid line). 
}
\label{fig:model2Gap+}
\end{figure}

Again estimation is made for two cases that (i) $\lambda^2 > 0$ and (ii)
$\lambda^2 < 0$.

\noindent (i) $\lambda^2 > 0$ : the solution of the gap equation (\ref{gapeq2v2}) 
can be categorized into three cases, (ia), (ib), and (ic), according to
the value of $\omega$: see the left of Fig.~\ref{fig:model2Gap+}.

\begin{itemize}
\item[(ia)] $\omega >0 $ : $\bar{\Omega}_0 = \omega$ then from Eq.~(\ref{OmegazeroGap})
$$
\mathrm{e}^{\bar{\Omega}_0 T} \stackrel{T \to \infty}{\longrightarrow} \infty \ ,
$$
so that all higher orders of $1/N$ in Eq.~(\ref{m2GroundStateFull}) do vanish, leaving the leading $O(N)$ term,
\begin{eqnarray}
E_{\mbox{\scriptsize{G}}}^{N} = E_{\mbox{\scriptsize{G}}}^{1} = E_{\mbox{\scriptsize{G}}}^{1/N} = 0 \ , 
\end{eqnarray}
which matches with the exact energy~(\ref{ExactSOLmodel2lamda+omega+}) in the appendix.

\item [(ib)] $\omega <0$ and $\lambda^2 \geq  |\omega|$ : $\bar{\Omega}_0 =0$ then from Eq.~(\ref{OmegazeroGap}) 
$$
\mathrm{e}^{\bar{\Omega}_0 T} = -1+\frac{\lambda^2}{|\omega|} \ ,
$$
so that there remains only the leading term in Eq.~(\ref{m2GroundStateFull}) to give
\begin{eqnarray}
E_{\mbox{\scriptsize{G}}}^{N} = E_{\mbox{\scriptsize{G}}}^{1} = E_{\mbox{\scriptsize{G}}}^{1/N} = -N \frac{|\omega|^2}{2 \lambda^2} \ . 
\end{eqnarray}

\item[(ic)] $\omega <0$ and $0 < \lambda^2 < |\omega|$ case: 
$\bar{\Omega}_0 = -|\omega| + \lambda^2( < 0)$ then
$$
\mathrm{e}^{\bar{\Omega}_0 T} \stackrel{T \to \infty}{\longrightarrow} 0 \ ,
$$
so that the leading term only gives us 
\begin{eqnarray}
E_{\mbox{\scriptsize{G}}}^{N} = E_{\mbox{\scriptsize{G}}}^{1} = E_{\mbox{\scriptsize{G}}}^{1/N} = N \left(- |\omega| +
				      \frac{\lambda^2}{2} \right) \ . 
\end{eqnarray}

\end{itemize}
We plot the result in the right of Fig.~\ref{fig:model2Gap+} from which we see that the leading approximation almost matches the exact curve. The deviation comes from the boundary of $\lambda^2$ domains, where discontinuity in the exact energy curve emerges. Note that the number of domains is three in the energy eigenvalue (generally $N+1$: see Eq.~(\ref{m2Lambda+omega-GroundEnergy2}) in the appendix) but two in AFM.

\begin{figure}[ht]
\begin{center}
\raisebox{0cm}{}
\epsfxsize=7cm \epsfbox{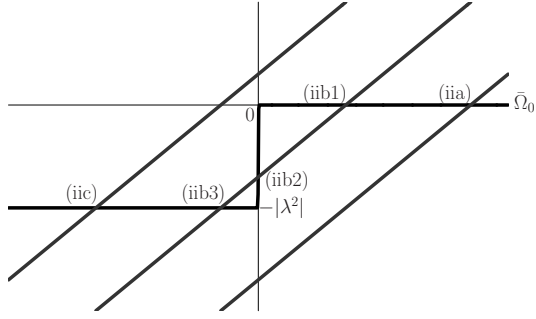} 
\end{center}
 \caption{The gap equation, $\lambda^2 < 0 $: (iia) ; $\omega > 0$
 and $ 0 < |\lambda^2| < \omega$. (iib) ; $\omega > 0$ and 
$|\lambda^2| \geq \omega$, where three solutions, 
(iib1) $\sim$ (iib3) , exist. (iic) ; $\omega < 0$.  
}
\label{fig:model2Gap-}
\end{figure}

\noindent (ii) $\lambda^2 < 0$. Five different solutions of Eq.~(\ref{gapeq2v2}) are found according the value of $\omega$. See Fig.~\ref{fig:model2Gap-}.

\begin{itemize}
\item[(iia)] $\omega > 0$ and $ 0 < |\lambda^2| < \omega$ : $\bar{\Omega}_0 = \omega$ then
$$
\mathrm{e}^{\bar{\Omega}_0 T} \stackrel{T \to \infty}{\longrightarrow} \infty  \ ; 
$$
so that the leading term in Eq.~(\ref{m2GroundStateFull}) only gives us 
\begin{eqnarray}
E_{\mbox{\scriptsize{G}}}^{N} = E_{\mbox{\scriptsize{G}}}^{1} = E_{\mbox{\scriptsize{G}}}^{1/N} = 0 \ . 
\end{eqnarray}

\item[(iib)] $\omega > 0$ and $|\lambda^2| \geq \omega$ : 
\begin{itemize}
 \item[(iib1)] $\bar{\Omega}_0 = \omega$ then
$$
{\mathrm e}^{\bar{\Omega}_0 T} \stackrel{T \to \infty}{\longrightarrow} \infty \ ; 
$$
so that again the leading term gives us
\begin{eqnarray}
E_{\mbox{\scriptsize{G}}}^{N} = E_{\mbox{\scriptsize{G}}}^{1} = E_{\mbox{\scriptsize{G}}}^{1/N} = 0 \ .  
\end{eqnarray}

\item[(iib2)] $\bar{\Omega}_0 = 0$ then from Eq.~(\ref{OmegazeroGap})
$$
{\mathrm e}^{\bar{\Omega}_0 T} = -1+\frac{|\lambda^2|}{\omega} \ ; 
$$
so that there is no correction other than the leading term yielding to
\begin{eqnarray}
E_{\mbox{\scriptsize{G}}}^{N} = E_{\mbox{\scriptsize{G}}}^{1}  = E_{\mbox{\scriptsize{G}}}^{1/N} = N \frac{\omega^2}{2 |\lambda^2|} \ ,
\end{eqnarray}
which is positive to be greater than (iib1) and should be discarded.

 \item[(iib3)] $\bar{\Omega}_0 = \omega - |\lambda^2| (< 0)$ then 
$$
 {\mathrm e}^{\bar{\Omega}_0 T} \stackrel{T \to \infty}{\longrightarrow} 0 \ , 
$$
so that there remains only the leading term in Eq.~(\ref{m2GroundStateFull}) to give
\begin{eqnarray}
E_{\mbox{\scriptsize{G}}}^{N} = E_{\mbox{\scriptsize{G}}}^{1}  = E_{\mbox{\scriptsize{G}}}^{1/N}  =  N \left( \omega -
				       \frac{|\lambda^2|}{2} \right)  = \left\{
                                                                                  \begin{array}{cc}
                                                                             \mbox{positive :} & \displaystyle{0 < |\lambda^2| < 2 \omega}   	\\
                                                                             \noalign{\vspace{2mm}}
\mbox{negative :}  &   \displaystyle{|\lambda^2| \geq 2 \omega}                                                                                  
                                                                                  \end{array}
                                                                                  \right.\ . \label{m2AFMlamda-omega+}
\end{eqnarray}  

\end{itemize}
Therefore we should adopt the solution (iia) up to $|\lambda^2 | = 2 \omega $ then switch it to (iib3): 
\begin{eqnarray}
 E_{\mbox{\scriptsize{G}}}  = \left\{
	 \begin{array}{ccc} 
	\displaystyle{N \left( \omega - \frac{|\lambda^2|}{2} \right)}  
	 & :  &   
\displaystyle{ |\lambda^2| \geq 2 \omega}  \\
	  \noalign{\vspace{2mm}}
	\displaystyle{0}   
	 & :  &  
	   \displaystyle{0 < |\lambda^2| < 2 \omega}  
	 \end{array}\right. \ ,   
\end{eqnarray}
which is nothing but the exact energy (\ref{m2ExactSOLlambda-omega+}) in the appendix.

\item[(iic)] $\omega < 0$ : $\bar{\Omega}_0 = -|\omega| - |\lambda^2|$ then 
$$
{\mathrm e}^{\bar{\Omega}_0 T} \stackrel{T \to \infty}{\longrightarrow} 0 \ , 
$$
so that there is no correction from the higher orders to obtain
\begin{eqnarray}
  E_{\mbox{\scriptsize{G}}}^{N} = E_{\mbox{\scriptsize{G}}}^{1} = E_{\mbox{\scriptsize{G}}}^{1/N}  = - N \left( |\omega| +  \frac{|\lambda^2|}{2}\right) \ , \label{m2AFMlambda-omega-}
\end{eqnarray} 
matching with the exact energy (\ref{m2ExactSOLlambda-omega-}) in the appendix.
\end{itemize}

We have checked the model up to the $O\! \left(1/N^2  \right) $ to find
that all higher order corrections vanish as well as that the leading
term reproduces the exact value except the case $\lambda^2 >0$ ; $\omega
<0 $. Therefore we can conclude that the model belongs to the WKB exact
class~\cite{rf:FFKS}, although the expansion is performed with respect
to $1/N$ which differs from the loop expansion(= WKB) in this case. A
slight discrepancy emerges from the boundaries where discontinuity is
eminent (The right of Fig.~\ref{fig:model2Gap+}.). 
As was stated before AFM is analytic so that if the number of the domains in $\lambda^2$ differs each other a slight deviation could inevitably occur.

\section{Discussion}
In this paper, we discuss validity of AFM in terms of quantum mechanical four-fermi models. The model with the anti-normal ordered form shows us that when $\lambda^2 > 0 \ ; \ \omega >0 $ the two-loop result almost fits the exact value (Fig.~\ref{fig:model1omega+lambda+}), when $\lambda^2 > 0 \ ; \ \omega < 0 $ the tree result becomes exact, and when $\lambda^2 < 0 \ ; \ \omega >0 $ (Fig.~\ref{fig:model1omega+lambda-}(left)) as well as when $\lambda^2 < 0 \ ; \ \omega <0 $ (Fig.~\ref{fig:model1omega+lambda-}(right)) the one-loop correction reproduces the exact energy. Even in $\lambda^2 > 0 \ ; \ \omega >0 $, we should expect the exact fit of the one-loop or tree result, but the non-analytic structure of the exact energy curve causes the deviation. 

In the second model, all the higher order corrections vanish other than the lowest one, showing us another example of the WKB exact class: when $\lambda^2 > 0 \ ; \ \omega >0 $ the leading term reproduces the exact value, when $\lambda^2 > 0 \ ; \ \omega < 0 $ it fits well except the region where non-analytic structure becomes eminent (Fig.~\ref{fig:model2Gap+}(right)), and when $\lambda^2 < 0 \ ; \ \omega >0 $ (Eq.~(\ref{m2AFMlamda-omega+})) as well as when $\lambda^2 < 0 \ ; \ \omega < 0 $ (Eq.~(\ref{m2AFMlambda-omega-})) it fits exactly. The only deviation in (Fig.~\ref{fig:model2Gap+}(right)) comes from the non-analytic structure of the exact energy curve.

\vspace{5mm}

So far discussions are made under the discrete time path integral representation which keep $N_t$ finite until the end of calculations, but in perturbation or WKB approximation people rely on the continuous time path integral which takes $N_t \mapsto \infty $ first. Of course, it is simpler and easier, but in some cases~\cite{rf:ST} we are forced to use the discrete time representation. Here we give another example: the model-(1) with $N_t \mapsto \infty$. The partition function (\ref{PF}) becomes
\begin{eqnarray}
 Z(T) &=& \int \int \mathcal{D}^N \xi \mathcal{D}^N \xi^*   
  \exp \biggl[  -\int_0^T d t 
 \biggl\{  (\mbit{\xi}^*(t) \cdot \dot{\mbit{\xi}}(t)) \nonumber \\ 
 && +\omega (\mbit{\xi}^*(t) \cdot \mbit{\xi}(t)) 
  +\frac{\lambda^2}{2 N}(\mbit{\xi}^*(t) \cdot \mbit{\xi}(t))^2 
  \biggr\} \biggr]_{\mbox{\scriptsize{AP}}} \ , 
\label{ContinuumPartitionFunction}
\end{eqnarray}  
where AP implies $\mbit{\xi }(0) = - \mbit{\xi }(T)$ . After introducing the auxiliary field $\sigma(t)$
it gives
\begin{eqnarray}
 Z(T) &=& \int \mathcal{D} \sigma \exp [-N I[\sigma]] \ ,
\label{continuumauxiliaryPartitionfunction}
\end{eqnarray}
where
\begin{eqnarray}
 I[\sigma] = \int_0^T d t \frac{\sigma^2(t)}{2} 
- {\rm Tr }\ln \left( \frac{d}{d t} + \omega + i \lambda \sigma (t) \right)
\ , 
\end{eqnarray}
with Tr being the functional trace. Find the constant solution $\sigma(t) = \bar{\sigma}_0$, yielding the gap equation
\begin{eqnarray}
 \frac{\delta I[\sigma]}{\delta \sigma (t)} \bigg|_{\bar{\sigma}_0}
= \bar{\sigma}_0-i \lambda \bar{S}(t,t) = 0 \ , \label{continuumGapeq}
\end{eqnarray}
and the Fermion propagator, 
\begin{eqnarray}
 \bar{S}(t,t') 
 &=& \frac{1}{2 \cosh \frac{\bar{\Omega}T}{2}}
  \left[ \theta(t-t') {\mathrm e}^{\bar{\Omega} \left\{ \frac{T}{2}-(t-t') \right\}} 
   -\theta(t'-t) {\mathrm e}^{-\bar{\Omega} \left\{ \frac{T}{2}+(t-t') \right\}}
  \right] \ , \label{continuumGreenfunction}
\end{eqnarray}
with
\begin{eqnarray}
 \bar{\Omega} \equiv \omega + i \lambda \bar{\sigma}_0  \ .
\end{eqnarray}
By noting $\theta (0) = 1/2$, Eq.~(\ref{continuumGapeq}) reads
\begin{eqnarray}
 \bar{\Omega}-\omega 
= - \frac{\lambda^2}{2 } \tanh \frac{\bar{\Omega} T}{2} \ .
  \label{continuumGapeqversion2}
\end{eqnarray}
The propagator of the auxiliary field (\ref{mesonpro}) turns out to be, 
\begin{eqnarray}
 \bar{\Delta}(p) &=& 1 
  -\frac{T b}{1+T b} \delta_{p 0} \ , \quad p \equiv \frac{2m\pi }{T} \quad (m = 0, \pm 1 , \pm 2 ,\cdots ) \ ; 
\label{continuum}
\end{eqnarray}
with 
\begin{eqnarray}
b =  \frac{\lambda^2}{4} 
\left(1-\tanh^2 \frac{\bar{\Omega} T}{2} \right) \ .
\end{eqnarray}
The ground state energy up to the two-loop ( Eq.~(\ref{2LoopEnergyInModel1}) under the discrete time ) is therefore given as
\begin{eqnarray}
 E_{\mbox{\scriptsize{G}}}^{\mbox{\scriptsize{2-loop}}}  &=&  
  -\frac{N}{2 \lambda^2} (\bar{\Omega}-\omega)^2 
   - N \lim_{T \to \infty} \frac{1}{T}
   \ln \cosh \frac{\bar{\Omega} T}{2} \nonumber \\ 
 &&  +\frac{1}{2} \lim_{T \to \infty} \frac{1}{T}
   \ln \left[1 + \frac{T \lambda^2}{4} 
	\left( 1-\tanh^2 \frac{\bar{\Omega}T}{2} \right)
 \right] \nonumber \\
&& -\frac{1}{N} \lim_{T \to \infty} \bigg[
-\frac{\lambda^4 T}{64}
\left( 1-\tanh^2 \frac{\bar{\Omega}T}{2} \right)
\left( 1-3 \tanh^2 \frac{\bar{\Omega}T}{2} \right)
\nonumber \\
&& \times \left(1+\frac{\lambda^2 T}{4} 
\left( 1-\tanh^2 \frac{\bar{\Omega}T}{2} \right) 
\right)^{-1} 
\nonumber \\
&&+\frac{\lambda^6 T^2}{768}
\left( 1-\tanh^2 \frac{\bar{\Omega}T}{2} \right)^2
\left( 3-19 \tanh^2 \frac{\bar{\Omega}T}{2} \right)
\nonumber \\
&& \times \left(1+\frac{\lambda^2 T}{4} 
\left( 1-\tanh^2 \frac{\bar{\Omega}T}{2} \right) 
\right)^{-2} \nonumber \\
&&+\frac{5 \lambda^8 T^3}{1536}
\tanh^2 \frac{\bar{\Omega}T}{2} 
\left( 1-\tanh^2 \frac{\bar{\Omega}T}{2} \right)^3
\nonumber \\
&& \times \left(1+\frac{\lambda^2 T}{4} 
\left( 1-\tanh^2 \frac{\bar{\Omega}T}{2} \right) 
\right)^{-3} 
\bigg] \ .
   \label{continuum2loop}  
\end{eqnarray}
\begin{figure}[ht]
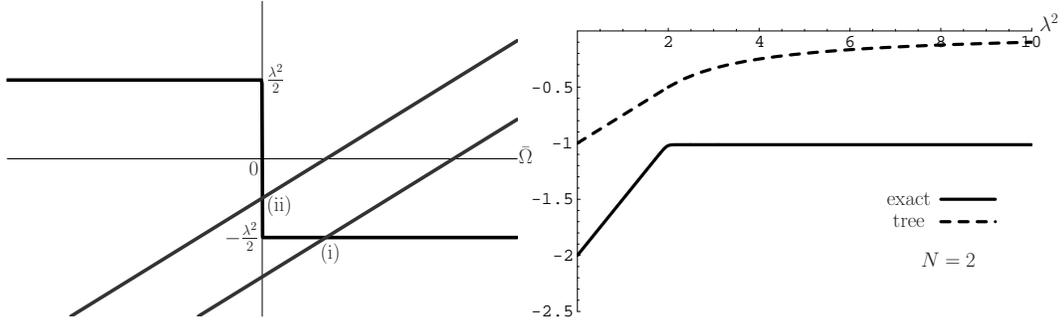

  \begin{center}
   \raisebox{0cm}{}
  \epsfxsize=7cm  \epsfbox{continuumGap.epsi} \epsfxsize=7cm  \epsfbox{continuumresult.epsi} 
  \end{center}
\caption{Left: the gap equation in continuum. \hfill  \hfill \hfill 
Right: the ground state energy for $N=2$ ($\lambda^2 > 0$ ; $\omega = 1$). The exact curve and the approximation result (only the lowest is left intact) are given by the solid and the dotted lines, respectively. Disagreement is prominent.}
\label{fig:continuumGap}
\end{figure}
We estimate the case $\lambda^2 > 0$ and $\omega > 0$. The solution of the gap equation~(\ref{continuumGapeqversion2}) can be categorized into two cases, (i) and (ii), according to the value of $\omega$: see Fig.~\ref{fig:continuumGap}.
\begin{itemize}
 \item[(i)] $\omega > 0$ and $ 0< \lambda^2 < 2 \omega$ : 
$\displaystyle{\bar{\Omega} = \omega-\frac{\lambda^2}{2}}$, then
$$
\tanh \frac{\bar{\Omega} T}{2}
\stackrel{T \to \infty}{\longrightarrow} -1 \ ,
$$ 
so that
\begin{eqnarray}
&&E_{\mbox{\scriptsize{G}}}^{\mbox{\scriptsize{tree}}} 
= E_{\mbox{\scriptsize{G}}}^{\mbox{\scriptsize{1-loop}}} 
= E_{\mbox{\scriptsize{G}}}^{\mbox{\scriptsize{2-loop}}} 
= N \left( \frac{\lambda^2}{8} - \frac{\omega}{2} \right)
\end{eqnarray}
 \item[(ii)] $\omega > 0$ and $\lambda^2 \geq 2 \omega$ :
$\bar{\Omega} = 0$, then
$$
\tanh \frac{\bar{\Omega} T}{2}
\stackrel{T \to \infty}{\longrightarrow} 
-\frac{2 \omega}{\lambda^2} \ ,
$$
so that 
\begin{eqnarray}
&&E_{\mbox{\scriptsize{G}}}^{\mbox{\scriptsize{tree}}} 
= E_{\mbox{\scriptsize{G}}}^{\mbox{\scriptsize{1-loop}}} 
= E_{\mbox{\scriptsize{G}}}^{\mbox{\scriptsize{2-loop}}} 
= - \frac{N \omega^2}{2 \lambda^2}
\end{eqnarray}
\end{itemize}
In continuum all the higher order corrections vanish. The result for $N=2$ and $\omega = 1$ is plotted in Fig.~\ref{fig:continuumGap}: disagreement with the exact value is apparent. In the discrete time (in Fig.~\ref{fig:model1omega+lambda+}), the tree result is improved by the higher loop effects but there is no higher loops.

\vspace{5mm}

There is a long history of studies in chiral symmetry breaking in 
QED~\cite{rf:CSBinQED}: in a strong coupling region there seems to exist a chiral breaking phase. However trials, such as the Schwinger-Dyson or the effective potential approach, have been suffered from gauge dependence. Here is an alternative: a starting Lagrangian of massless QED reads,
\begin{eqnarray}
{\cal L} & = & - \overline{\psi}\gamma_\mu \left( \partial_\mu - ie A_\mu  \right) \psi - \frac{1}{4} F_{\mu \nu}^2 \ ; \qquad    F_{\mu \nu} \equiv \partial_\mu A_\nu -  \partial_\nu A_\mu \ ; 
\end{eqnarray}
so that the partition function,
\begin{eqnarray}
Z = \int {\cal D} A_\mu {\cal D} \psi {\cal D} \overline{\psi} \exp  \left( \int d^4 x  {\cal L} \right)    \ ; 
\end{eqnarray}
can be, with the aid of the auxiliary fields, reduced to 
\begin{eqnarray}
Z & = & \int {\cal D} B_\mu {\cal D} C_\mu  {\cal D} \psi {\cal D} \overline{\psi} \exp  \left( \int d^4 x  {\cal L'} \right)   \ ; \\
{\cal L'}& =& - \overline{\psi}\gamma_\mu \left( \partial_\mu - ie B_\mu + e \gamma_5 C_\mu  \right) \psi - \frac{\mathrm{e}^2}{2 m^2}\left\{  \left(  \overline{\psi} \psi  \right)^2 -  \left(  \overline{\psi}\gamma_5  \psi  \right)^2  \right\}  \nonumber \\
& & - \frac{m^2}{2} B_\mu \left\{ \frac{- \dbox \delta_{\mu \nu }+ \partial_\mu \partial_\nu}{-\dbox + m^2} \right\}B_\nu  - \frac{m^2}{2}C_\mu^2 \ ;  \quad \left( \dbox \equiv \partial_\mu^2  \right) 
\end{eqnarray}
where use has been made of the Fierz identity. Note that there is no gauge fixing, the remnant of which can be seen from the invariance, $B_\mu \mapsto B_\mu + \partial_\mu \Lambda $. The four-fermi form looks similar to the Nambu--Jona-Lasinio model so that the chiral structure that whether the quantity, $\langle \overline{\psi } \psi \rangle $, is zero or nonzero would be explored in a parallel manner. This is our next task.

\vspace{10mm}

\noindent 
This work is supported by Grant-in-Aid for Science Research 
from the Japan Ministry of Education, Science and Culture; 13135217.

\appendix

\section{The exact ground state energy}
In this appendix calculations are made for the ground state energy (\ref{GroundStateEnergy}).

\subsection{Model-(1)}
First complete the square in the energy eigenvalue (\ref{Eeq}) to give 
\begin{eqnarray}
E_n = \frac{\lambda^2}{2 N} \left\{ n - \left(N - \frac{1}{2}
- \frac{N \omega}{\lambda^2} \right) \right\}^2 
- \frac{N}{2 \lambda^2} \left( \omega + \frac{\lambda^2}{2 N} \right) \ . 
\label{squaredEeq}
\end{eqnarray}
The vertex is thus 
\begin{eqnarray}
n_{\mbox{\scriptsize vert}}= N-\frac{1}{2}- \frac{N \omega}{\lambda^2}
 \ ,
\end{eqnarray}
where the minimum or maximum occurs if there is no restriction for $n$. However $ 0 \leq n \leq N$ so we need a detailed inspection.

\begin{itemize}
\item[(i)] $\lambda^2>0$ and $\omega > 0 $ : $E_n$ is concave and $n_{\mbox{\scriptsize vert}} < \displaystyle{\frac{2N-1}{2}}$. The pattern of $E_n$ is given in Fig.~\ref{fig:concavepattern}. From Fig.~\ref{fig:concavepattern}(a), if
$$
n_{\mbox{\scriptsize vert}} < \frac{1}{2} \Leftrightarrow 0< \lambda^2 < \omega \left(\frac{N}{N-1} \right)\ ; 
$$ 
$n=0$ is the minimum to give
\begin{eqnarray*}
E_{\mbox{\scriptsize{G}}} = E_0 = \frac{N-1}{2}\lambda^2-N \omega  \ .
\end{eqnarray*}
\begin{figure}[ht]
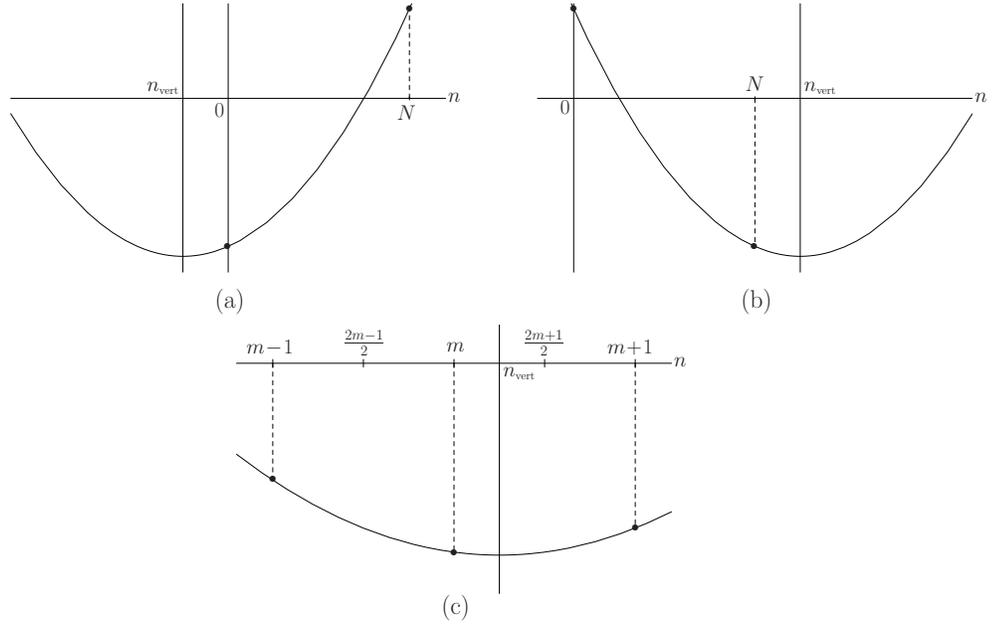

\begin{eqnarray*}
& & \epsfxsize=6cm \epsfbox{concavecase1.epsi}  \hspace{10mm}
\epsfxsize=6cm \epsfbox{concavecase2.epsi}  \\
& & \hspace{30mm}\epsfxsize=6cm \epsfbox{concavecase3.epsi}
\end{eqnarray*}  
\caption{The patterns of the ground state energy for $\lambda^2 > 0$. 
(a) shows the case of 
$n_{\mbox{\scriptsize vert}} <  \displaystyle{\frac{1}{2}}$. The minimum
 occurs at $n=0$. (b) shows the case of 
$n_{\mbox{\scriptsize vert}} \geq \displaystyle{\frac{2N-1}{2}}$. The
 minimum occurs at $n=N$. (c) shows the case of
 $\displaystyle{\frac{1}{2}} \leq n_{\mbox{\scriptsize vert}} < \displaystyle{\frac{2N-1}{2}}$. 
When $\displaystyle{\frac{2m-1}{2}} \leq
 n_{\mbox{\scriptsize vert}} < \displaystyle{\frac{2m+1}{2}} \ ; (m=
 1, 2, \dots , N-1)$ the minimum occurs at $n=m$. 
}
\label{fig:concavepattern}
\end{figure}
Generally, from Fig.~\ref{fig:concavepattern}(c), if
$$
\frac{2m-1}{2} \leq n_{\mbox{\scriptsize vert}} < \frac{2m+1}{2} \Leftrightarrow \omega \left(\frac{N}{N-m} \right) \leq \lambda^2 < \omega \left(\frac{N}{N-m-1} \right) \ ; 
$$
$n=m (1, 2, \dots  N-1)$ is the minimum to give 
\begin{eqnarray*}
E_{\mbox{\scriptsize{G}}} = E_m = \frac{(N-m)(N-m-1)}{2 N}\lambda^2  -(N-m) \omega  \ .
\end{eqnarray*}
Cases can be summarized to give
\begin{eqnarray}
\omega \left(\frac{N}{N-m} \right) \leq \lambda^2 < \omega \left(\frac{N}{N-m-1} \right)  \ ,  \quad (m= 0 , 1, 2, \dots N-1) \ ; \label{lambda+omega+Domain}
\end{eqnarray}
then
\begin{eqnarray}
 E_{\mbox{\scriptsize{G}}}  = E_m = \frac{(N-m)(N-m-1)}{2 N}\lambda^2  -(N-m) \omega  \ .       
  \label{exa1}
\end{eqnarray}  
(Note that when $m=0$ the lower bound $\omega$ of $\lambda^2$ should be discarded.) As $N$ goes large the number of domains given by Eq.~(\ref{lambda+omega+Domain}) increases, which smoothes the discontinuity of the energy curve around the boundaries as seen from Fig.~\ref{fig:N2 5 10Tree} with $N=2, 5$ and $10$.

\item[(ii)] $\lambda^2 > 0$ and $\omega < 0$ : 
$$
n_{\mbox{\scriptsize vert}} = N - \frac{1}{2} + \frac{N |\omega| }{\lambda^2 } > N - \frac{1}{2} \ . 
$$
From Fig.~\ref{fig:concavepattern}(b), $n = N$ is the lowest. $E_N = 0$ from Eq.~(\ref{squaredEeq}), so that
\begin{eqnarray}
 E_{\mbox{\scriptsize{G}}} = E_N = 0 \ . \label{ExactSOLlambda+omega-}
\end{eqnarray} 

\item[(iii)] $\lambda^2 < 0$ and $\omega > 0$ : $E_n$ is convex and the pattern of $E_n$ is shown in Fig.~\ref{fig:convexpattern}.
$$
n_{\mbox{\scriptsize vert}} = N - \frac{1}{2} +  \frac{N \omega }{|\lambda^2| } > \frac{N}{2}  \ ; 
$$
since $N\geq 2$. Therefore from Fig.~\ref{fig:convexpattern}(b), $n=0$ is the minimum and $\displaystyle{E_0 =  -\frac{N-1}{2}|\lambda^2|-N \omega}$ so that
 \begin{eqnarray}
 E_{\mbox{\scriptsize{G}}} = E_0 =  -\frac{N-1}{2}|\lambda^2|-N \omega \ . \label{ExactSOLlambda-omega+}
\end{eqnarray}

\begin{figure}[ht]
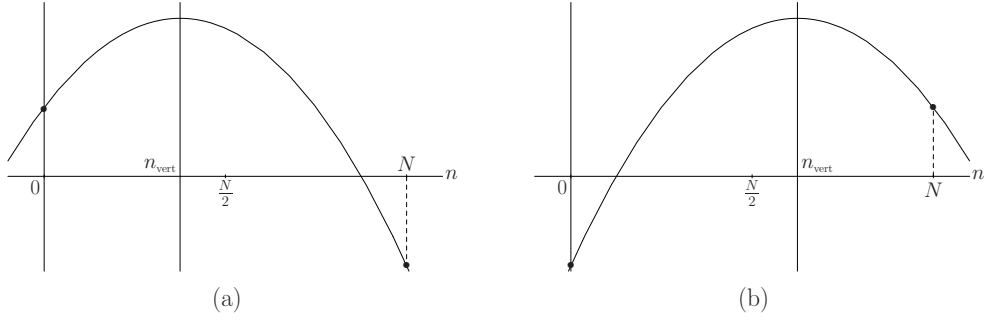


\begin{eqnarray*}
\epsfxsize=6cm \epsfbox{convexcase1.epsi}  \hspace{10mm}
\epsfxsize=6cm \epsfbox{convexcase2.epsi}   
\end{eqnarray*}  
\caption{The patterns of the ground state energy for $\lambda^2 < 0$. 
(a) shows the case of $\displaystyle{n_{\mbox{\scriptsize vert}} \leq \frac{N}{2}}$. The minimum occurs at $n=N$. (b): the case of $\displaystyle{\frac{N}{2} < n_{\mbox{\scriptsize vert}}}$. The minimum occurs at $n=0$. 
}
\label{fig:convexpattern}
\end{figure}

\item[(iv)] $\lambda^2 < 0$ and $\omega < 0$ :
$$
n_{\mbox{\scriptsize vert}} = N - \frac{1}{2} -  \frac{N |\omega| }{|\lambda^2| }  \ . 
$$
From Fig.~\ref{fig:convexpattern}(a), if
\begin{eqnarray*}
n_{\mbox{\scriptsize vert}} \leq \frac{N}{2} \Leftrightarrow 0 < |\lambda^2| \leq |\omega| \frac{2N}{N-1} \ ; 
\end{eqnarray*}
$n=N$ is the minimum to give $E_N =0$. Further from Fig.~\ref{fig:convexpattern}(b), if
\begin{eqnarray*}
\frac{N}{2} < n_{\mbox{\scriptsize vert}}  \Leftrightarrow  |\omega| \frac{2N}{N-1} < |\lambda^2| \ ; 
\end{eqnarray*}
$n=0$ is the minimum to give $\displaystyle{E_0 =  -\frac{N-1}{2}|\lambda^2|+N |\omega|} $.
Therefore 
\begin{eqnarray}
 E_{\mbox{\scriptsize{G}}}  = \left\{
	 \begin{array}{ccc} 
	\displaystyle{ 0}  
	 & :  &   
\displaystyle{  0 < |\lambda^2| \leq   \frac{2N}{N-1} |\omega|}  \\
	  \noalign{\vspace{2mm}}
	\displaystyle{ -\frac{N-1}{2}|\lambda^2| + N |\omega|}  
	 & :  &  
	   \displaystyle{ |\lambda^2| > \frac{2N}{N-1} |\omega|}  
	 \end{array}\right. \ .   \label{ExactSOLlambda-omega-}
\end{eqnarray}

\end{itemize}

\subsection{Model-(2)}
Again complete the square in the energy eigenvalue~(\ref{Eeqv2}) to yield
\begin{eqnarray}
 E_n = \frac{\lambda^2}{2N} \left(n + \frac{N \omega}{\lambda^2} \right)^2
-\frac{N \omega^2}{2 \lambda^2} \ . \label{squaredEeqv2}
\end{eqnarray}
Then the vertex is 
\begin{eqnarray}
 n_{\mbox{\scriptsize vert}} = - \frac{N \omega}{\lambda^2} \ .
\end{eqnarray}

\begin{itemize}
\item[(i)] $\lambda^2 > 0$ and $\omega > 0$ : $E_n$ is concave. If
$$
n_{\mbox{\scriptsize vert}} = - \frac{N \omega}{\lambda^2} < 0 \ ; 
$$ 
from Fig.~\ref{fig:concavepattern}(a) $n=0$ is the minimum to give
\begin{eqnarray}
 E_{\mbox{\scriptsize{G}}} = E_0 = 0 \ . \label{ExactSOLmodel2lamda+omega+}
\end{eqnarray} 

\item[(ii)] $\lambda^2 > 0$ and $\omega < 0$ : 
$$
n_{\mbox{\scriptsize vert}} =  \frac{N |\omega|}{\lambda^2}  \ , 
$$
which is positive. From Fig.~\ref{fig:concavepattern}(a), if
\begin{eqnarray}
n_{\mbox{\scriptsize vert}} < \frac{1}{2} \Leftrightarrow   \lambda^2 > 2N |\omega | \ ; \label{m2Domain1/2}
\end{eqnarray}
the minimum occurs at $n=0$ to give
$$
E_{\mbox{\scriptsize{G}}} = E_0 = 0 \ . 
$$
Generally if
\begin{eqnarray}
\frac{2m-1}{2} \leq n_{\mbox{\scriptsize vert}} < \frac{2m+1}{2} \Leftrightarrow  \frac{2N}{2m+1}|\omega | <  \lambda^2 \leq  \frac{2N}{2m-1}  |\omega |\ ; \label{model2DomainInterval}
\end{eqnarray}
from Fig.~\ref{fig:concavepattern}(c), $n=m (m=1,2,\dots , N-1)$ is the minimum to give 
\begin{eqnarray}
 E_{\mbox{\scriptsize{G}}} = E_m = - |\omega| m + \frac{\lambda^2}{2N}m^2  \ . \label{m2Lambda+omega-GroundEnergy}
\end{eqnarray}
When $n_{\mbox{\scriptsize vert}} \geq  \displaystyle{\frac{2N-1}{2}}$ 
\begin{eqnarray}
0 < \lambda^2 \leq \frac{2N}{2N-1}|\omega|  \ ; \label{m2Domainm>N}
\end{eqnarray} 
from Fig.~\ref{fig:concavepattern}(b), the minimum occurs at $n=N$ to give
$$
E_{\mbox{\scriptsize{G}}} = E_N = -N |\omega|  + \frac{N \lambda^2}{2}   \ . 
$$

In view of Eqs.~(\ref{m2Domain1/2}), (\ref{model2DomainInterval}), and (\ref{m2Domainm>N}), we can take the ground state energy as Eq.~(\ref{m2Lambda+omega-GroundEnergy}) 
\begin{eqnarray}
 E_{\mbox{\scriptsize{G}}} = E_m = - |\omega| m + \frac{\lambda^2}{2N}m^2  \ ;  \quad  m = 0 , 1, \dots, N \ ; \label{m2Lambda+omega-GroundEnergy2}
\end{eqnarray}
in the region 
\begin{eqnarray}
 \frac{2N}{2m+1}|\omega | <  \lambda^2 \leq  \frac{2N}{2m-1}  |\omega | \ ; \quad m = 0 , 1, \dots, N \ ; \label{model2Lambda+omega-Domain}
\end{eqnarray}
under the condition  $\displaystyle{ \left. \frac{2N}{2m+1}|\omega |\right|_{m=N} \hspace{-6mm} \Longrightarrow  0} $ and $\displaystyle{ \left. \frac{2N}{2m-1}|\omega |\right|_{m=0} \hspace{-6mm} \Longrightarrow  \infty }$.

\item[(iii)] $\lambda^2 < 0$ and $\omega > 0$ : $E_n$ is convex.
$$
n_{\mbox{\scriptsize vert}} =  \frac{N \omega}{|\lambda^2|}  \ . 
$$
The situation is similar to (iv) in Model-(1). If
$$
n_{\mbox{\scriptsize vert}} \leq \frac{N}{2} \Leftrightarrow  |\lambda^2| \geq 2\omega  \ ; 
$$
then from Fig.~\ref{fig:convexpattern}(a), $n = N$ is the minimum to give 
$$
E_{\mbox{\scriptsize{G}}} = E_N = N \left( \omega - \frac{|\lambda^2|}{2} \right) \ . 
$$
Further if
$$
\frac{N}{2} < n_{\mbox{\scriptsize vert}}   \Leftrightarrow  0 <  |\lambda^2| < 2\omega  \ ; 
$$
from Fig.~\ref{fig:convexpattern}(b), $n=0$ is the minimum to give 
$$
E_{\mbox{\scriptsize{G}}} = E_0 =0 \ . 
$$
Therefore
\begin{eqnarray}
 E_{\mbox{\scriptsize{G}}}  = \left\{
	 \begin{array}{ccc} 
	\displaystyle{N \left( \omega - \frac{|\lambda^2|}{2} \right)}  
	 & :  &   
\displaystyle{|\lambda^2| \geq 2 \omega}  \\
	  \noalign{\vspace{2mm}}
	\displaystyle{0}   
	 & :  &  
	   \displaystyle{0 < |\lambda^2| <  2 \omega}  
	 \end{array}\right. \ .   \label{m2ExactSOLlambda-omega+}
\end{eqnarray}

\item[(iv)] $\lambda^2 < 0$ and $\omega < 0$ :
$$
n_{\mbox{\scriptsize vert}} =  - \frac{N |\omega|}{|\lambda^2|}  \ , 
$$
which is always negative so that from Fig.~\ref{fig:convexpattern}(a) $n=N$ is the minimum: 
\begin{eqnarray}
 E_{\mbox{\scriptsize{G}}} = E_N = -N \left( |\omega| + \frac{|\lambda^2|}{2} \right)   \ . \label{m2ExactSOLlambda-omega-}
\end{eqnarray} 

\end{itemize}

\end{document}